\documentclass[iop]{emulateapj}
\usepackage{amsmath,amssymb,times,graphics,morefloats}
\usepackage{gensymb}
 \usepackage{longtable}
 \usepackage{ tipa }
\usepackage{color}

\newcommand{\cwdm}{WD+dM\ }
\newcommand{\cwdms}{WD+dMs\ }

\defcitealias{K09}{K09}

\begin{document}

\title{Using Close White Dwarf + M Dwarf Stellar Pairs to Constrain the Flare Rates in Close Stellar Binaries}
\journalinfo{{\it The Astronomical Journal}, Accepted}
\submitted{Submitted: December 11, 2015; Accepted: February 17, 2016 }

\author{Dylan P. Morgan\altaffilmark{1,2}, Andrew A. West\altaffilmark{1}, Andrew C. Becker\altaffilmark{3}}

\altaffiltext{1}{Astronomy Department, Boston University, 725 Commonwealth
  Ave, Boston, MA 02215, USA}
\altaffiltext{2}{Corresponding author: dpmorg@bu.edu}
\altaffiltext{3}{Department of Astronomy, University of Washington, Box 351580, Seattle, WA 98195, USA}

\begin{abstract}
We present a study of the statistical flare rates of M dwarfs (dMs) with close white dwarf (WD) companions (WD+dM; typical separations $<$ 1 au). Our previous analysis demonstrated that dMs with close WD companions are more magnetically active than their field counterparts. One likely implication of having a close binary companion is increased stellar rotation through disk-disruption, tidal effects, and/or angular momentum exchange; increased stellar rotation has long been associated with an increase in stellar activity. Previous studies show a strong correlation between dMs that are magnetically active (showing H$\alpha$ in emission) and the frequency of stellar flare rates. We examine the difference between the flare rates observed in close \cwdm binary systems and field dMs. Our sample consists of a subset of 181 close \cwdm pairs from \cite{Morgan2012} observed in the Sloan Digital Sky Survey Stripe 82, where we obtain multi-epoch observations in the Sloan {\it ugriz}-bands. We find an increase in the overall flaring fraction in the close \cwdm pairs (0.09$\pm$0.03\%) compared to the field dMs \citep[0.0108$\pm$0.0007\%;][]{Kowalski2009} and a lower flaring fraction for active \cwdms (0.05$\pm$0.03\%) compared to active dMs \citep[0.28$\pm$0.05\%;][]{Kowalski2009}. We discuss how our results constrain both the single and binary dM flare rates. Our results also constrain dM multiplicity, our knowledge of the Galactic transient background, and may be important for the habitability of attending planets around dMs with close companions.

\end{abstract}


\section{Introduction}\label{Sec:Intro}
Despite M dwarfs (dMs) being upwards of four orders of magnitude less luminous than the Sun \citep{Bochanski2011}, they are capable of producing flares with 100-1000 times more energy than the largest Solar flares \citep{Hilton2011}. In dMs, these flares manifest themselves as an increase in optical and ultraviolet (UV) continuum emission equivalent to a 10,000 K blackbody \citep{Hawley1991, Kowalski2014}. The flares are thought to occur during magnetic reconnection events, which inject magnetic energy into the atmosphere of the dM and can be observed in the X-ray \citep[e.g.,][]{Osten2010}, UV \citep{Robinson2005, Hawley2007}, Optical \citep{Kowalski2009, Walkowicz2011}, Infrared \citep{Schmidt2012}, and Radio \citep{Stepanov2001, Osten2008}. The flare morphology has been categorized into two primary phases: 1) the impulsive phase, characterized by a quick increase in continuum emission within a few minutes or less \citep[e.g.,][]{Hawley1991, Eason1992, Kowalski2010}, and 2) a gradual decay phase, which can last from tens of minutes to a few hours \citep{Moffett1974, Zhilyaev2007}. The transient and stochastic nature of stellar flares have made them difficult and expensive to observe, traditionally requiring continuous time-resolved photometry of individual stars, both from the ground \citep{Moffett1974, Lacy1976} and from space \citep{Audard2000, Gudel2003}.  However, recent all-sky surveys like the Sloan Digital Sky Survey \citep[SDSS;][]{Abazajian2009} allow us to characterize the flaring properties of populations of dMs \citep[e.g.,][]{Kowalski2009, Hilton2010} instead of individual systems.

It has become increasingly important to know how frequently flares occur and the amount of energy released in dM flare events, information that can be gleaned from individual systems as well as for large ensembles of low-mass stars. High energy radiation and particles from flares and associated Coronal Mass Ejections \citep[CMEs; e.g.,][]{Aarnio2011} may have adverse effects on the habitability of exoplanets by altering the size and composition of atmospheres or by directly damaging DNA \citep[e.g.,][]{Khodachenko2007, Lammer2007, Scalo2007}. \cite{Dressing2013, Dressing2015a} used the Kepler Object of Interest (KOI) catalog to predict that every dM hosts at least one terrestrial planet with an orbital period less than 50 days, one in two dMs likely have Earth-sized planets (0.5-1.5 Earth radii), and one in seven dMs will have an Earth-sized planet in the habitable zone (HZ). Another study predicted that at least 50\% of dMs likely host at least 6 terrestrial planets \citep{Ballard2014}, a finding corroborated by the recent discovery of the closest Earth-analog exoplanet (1.2 Earth radii) discovered around an dM \citep[$\sim$ 0.21 M$_{\odot}$;][]{Berta-Thompson2015}. Considering that 70\% of all the stars in the Galaxy are dMs \citep[e.g.,][]{Henry1994, Chabrier2003, Reid2004, Winters2015}, together with the predicted frequency of habitable planets around dMs, the Transiting Exoplanet Survey Satellite \citep[TESS;][]{Ricker2014} will produce a statistically significant sample of nearby potentially habitable worlds. However, to properly address the habitability of these planets, it is crucial that we understand the flaring frequency, magnetic activity, and space weather environment of the host stars. 

Additionally, many of the short-lived optical transients in time-domain surveys are Galactic dM flares \citep[e.g,][]{Becker2004, Rykoff2005, Kulkarni2006, Rau2008, Berger2013}, and thus, contribute significant noise to the Galactic and extragalactic transient background. These transients are of particular importance for time-domain surveys such as Panoramic Survey Telescope \& Rapid Response System \citep[Pan-STARRS; ][]{Kaiser2004} and the upcoming Large Synoptic Survey Telescope \citep[LSST Science Collaborations;][]{LSST2009}, which are interested in extragalactic transient phenomena such as Novae (normal, sub-luminous, and super-), high-energy transients (gamma-ray bursts and X-ray flashes), as well as degenerate binary mergers \citep{LSST2009}. M dwarf flares are much more common then these exotic transients \citep{Berger2013} and can masquerade as optical transients, leading to erroneous and expensive spectroscopic follow-up. This is particularly important for LSST, where \cite{Hilton2011} predicted that each full-frame LSST image will have an dM flare and, furthermore, 1\% of all LSST frames will have an dM flare appearing as an optical transient; whereas normally, the faint dM would be below the detection threshold in quiescence.

Having a proper understanding of the cumulative number of flares as a function of both energy and time, or what is known as the Flare Frequency Distribution (FFD), is important for constraining how dM flares affect the habitability of attending planets as well as their contribution to the Galactic transient background. Progress in understanding the dM FFD began with extensive ground-based observations of eight active dM stars \citep[over 400 hours spanning 2 years;][]{Moffett1974,Lacy1976}, in which late-type (M6-M9) dMs were found to flare more frequently, but at lower energies than early-type (M0-M3) dMs. This trend has been attributed to selection effects, in which low-energy flares are both more common than high-energy flares and easier to observe around less luminous, late-type dMs. Similar, but more recent, ground-based monitoring of eight dMs found that flares are also observed in inactive early-type dMs, albeit at a much lower frequency than the active dMs \citep{Hilton2011}. With the advent of large all-sky surveys like SDSS, these studies can be expanded to large ensembles of dMs. \citet[][hereafter K09]{Kowalski2009} characterized the flare properties of populations of dMs utilizing the SDSS Stripe 82 (S82) catalog, Stripe 82 a small region of sky observed multiple times by SDSS over a ten year baseline. K09 reported 271 flares across 2.5 million individual observations, estimating an dM Galactic flare rate of 1.3 flares hr$^{-1}$ deg$^{-2}$ (for flares $ \Delta {\it u} \ge 0.7$ and ${\it u} < 22$). 

One parameter that is still not well constrained is the effect of binarity on dM flares, specifically, close binarity ($<$ 100 au). M dwarf multiplicity represents a non-negligible fraction of the dM population, predicted to be anywhere from 12-27\% \citep{Janson2012, Janson2014, Winters2015}. A recent study showed that dMs with close stellar companions are more likely to be magnetically active (as measured by H$\alpha$) and remain active for longer than isolated field dMs \citep{Morgan2012}. Stars in close binary systems ($<$ 100 au) likely have short-lived disks and thus are unable to properly dissipate angular momentum through mechanisms like magnetic disk locking \citep[e.g.,][]{Koenigl1991,Shu1994} or accretion powered stellar winds \citep[][]{Korycansky1995, Papaloizou1997, Matt2012}, resulting in faster rotation than if they had evolved as a single star. The closest binary pairs (separations $<$10 au) will also undergo the effects of tidal synchronization and over time will synchronize their spin rate to their orbital motion, again, resulting in faster than normal rotation \cite[e.g.,][]{Meibom2005}. Unfortunately, building a significant sample of close binary systems with an dM component can be challenging. Equal mass binaries are difficult to distinguish from low-resolution spectroscopic observations aside from the very closest pairs (which will have radial velocity broadened spectroscopic features), and a higher mass companion will be much more luminous, obscuring the dM from view. However, white dwarf + dM (WD+dM) pairs are ideal for identifying close, unresolved binaries due to their similar luminosities and different colors. Their properties allow them to be selected from photometric catalogs with high-fidelity \citep[84\%;][]{Rebassa-Mansergas2013} and can be separated into their individual components from low-resolution spectra. Thus, a sample of unresolved \cwdm pairs provides the perfect laboratory for testing how close binarity affects flaring of dMs.

We extend the study of magnetic activity around \cwdms to include flares around close \cwdm pairs using the SDSS S82 time-domain catalog. In accordance with \cwdms having a higher fraction of magnetic activity \citep[as measured by H$\alpha$;][]{Morgan2012}, we expect to find a higher fraction of flares around \cwdms when comparing to the field dM flare study from SDSS S82 (K09). We carefully prepare a \cwdm sample from S82 to ensure it is comparable to the field dM sample. Our spectroscopic sample makes use of a subset of previous SDSS spectroscopic \cwdm samples \citep{Morgan2012,Rebassa-Mansergas2012,Rebassa-Mansergas2013} that are matched to the S82 database. Of particular interest is determining the binary dM vs. isolated dM flare frequency as a function of spectral type and flare energy by using \cwdm flare rates as a proxy for binary dM flare rates. If the \cwdms flare more frequently and at higher energies than the field dM sample, it could have important implications for the potential habitability of circumbinary exoplanets \citep{Doyle2011, Welsh2012, Orosz2012, Kostov2013}. Increases in magnetic activity and flaring may not be isolated to stellar binaries; preliminary evidence suggests that dMs with giant close-in planets can also cause an increase magnetic activity \citep[seen in the X-ray;][]{Poppenhaeger2011,Poppenhaeger2014}. In the context of the Galactic transient background, if the \cwdm flare frequency is appreciably high, time-domain surveys may need to take into account close binary systems as they have colors similar to extragalactic sources (e.g., QSOs), rather than colors typical of field dMs.

In this paper, we use a large sample of both spectroscopic and photometric \cwdms to extend the \cite{Morgan2012} study of magnetic activity to include flares. In Section~2, we describe the data from the SDSS and SDSS S82 as well as the spectroscopic and photometric sample selection. In Section~3, we describe our procedure for selecting flares and the properties of the flaring sample. In Section~4, we compare our flaring sample to that of the field dM sample (K09). Finally, we discuss and summarize our findings in Section~5.

\section{Data}\label{Sec:Data}
This study uses data from the Sloan Digital Sky Survey \citep{York2000, Stoughton2002, 
Pier2003, Ivezic2004}, a large multicolor photometric \citep{Gunn1998,Doi2010} and spectroscopic \citep{Smee2013} survey covering 14,000 deg$^{2}$ centered on the Northern Galactic Cap along with three supplemental imaging strips in the Southern Galactic Cap \citep{Abazajian2009}. The 2.5m SDSS telescope \citep{Gunn2006} operates in drift-scan mode and observes to a depth of 22.0, 22.2, 22.2, 21.3, 20.5 magnitudes in the five filters {\it u, g, r, i ,z}, respectively. As of July, 2014, with Data Release 12 \citep[DR12;][]{Alam2015}, SDSS has observed $\sim$500 million unique photometric sources along with $>$4 million spectra.

Transient phenomena were not originally part of the science goals for SDSS. However, SDSS took repeat observations of a region of sky during Fall months when the Northern Galactic Cap fields were near the horizon. S82, is a small ($\sim$275 deg$^2$) region along the celestial equator in the Southern Galaxy ($-51\degree < \alpha < +60\degree, -1.266\degree <  \delta < +1.266\degree$) with over 20 observations in each filter spanning from 2001-2007 \citep{Abazajian2009}. S82 was originally designed as a testbed for probing deeper magnitudes by co-adding multiple observations, however, S82 has been used for numerous transient studies: e.g., supernova surveys \citep{Frieman2008, Sako2008, Kostrzewa-Rutkowska2013}, quasar variability \citep{Palanque-Delabrouille2011,Gu2011}, stellar variability \citep{Watkins2009, Sesar2010, Suveges2012} and flare rates in dMs (K09). In this paper, we use S82 data to understand the flare rates around close \cwdm binary pairs using a similar approach to that taken by K09.

\subsection{Spectroscopic Sample}
Our initial sample of 1756 spectroscopic \cwdm pairs from SDSS Data Release 10, was taken from \cite{Morgan2012}. We use the same color-cut methods as described in \cite{Morgan2012} to search for additional \cwdm pairs in SDSS DR12, and we find a total of 1933 spectroscopically confirmed \cwdm pairs. 

Using the procedures outlined in \cite{Morgan2012}, we adopt an iterative process for separating the WD and dM spectral components using WD models \citep{Koester2001} and high signal-to-noise dM templates \citep{Bochanski2010}. Each spectral component is fit individually; in subsequent iterations, the best fit model or template is subtracted from the original spectrum, and the fitting procedure is repeated. We repeat this process for ten iterations, or until we converge on a consistent solution for the dM and WD components. Typically our procedure converges in fewer than five iterations. Through this fitting procedure, we are able to measure the dM spectral types from the best fit dM templates and the WD spectral type (DA/DB), effective temperature (T$_{eff}$), and surface gravity ($\log(g)$) from the best fit WD models. Additionally, we measure the following parameters: dM magnetic activity through visual inspection of H$\alpha$ emission, H$\alpha$ equivalent width using a trapezoidal integration technique centered on the H$\alpha$ line (refer to \citet{West2004} and \citet{Morgan2012} for more details), distances using spectroscopic parallaxes \citep{Bochanski2008}, and absolute height above the Galactic plane.

Using the radial velocity (RV) shifts in the individual dM and WD spectral components, we estimate the projected physical separations. The RVs are calculated using a cross-correlation technique with the best fit (dM) or model (WD) as a reference rest template. All SDSS spectra are composite spectra made up of multiple exposures (at least 3 but as many as 15). The typical exposure times were roughly 900$-$1200s, and were often made in succession. The composite spectra, in addition to the repeat spectroscopic observations, allows for time-variability of the RV measurements in our systems. The separations are calculated assuming a Keplerian circular orbit, edge-on inclination and using measured RVs from the WD and dM spectral components (we use multi-epoch RVs when multiple SDSS spectra were available). We balance the centripetal and gravitational forces to find individual component separations from the center-of-mass of the system. 

We assign a mass of 0.6 M$_{\odot}$ to the WDs, which is empirically found to be the average mass of 90\% of WDs in SDSS Data Release 4 \citep{Kepler2007}. We find a similar average WD mass of $\sim$0.6 M$_{\odot}$ from a subset of \cwdms from \cite{Morgan2012} that have precise WD mass measurements from WD Cooling analysis \citep[e.g.,][]{Catalan2008}, confirming that assigning a mass of 0.6 M$_{\odot}$ for WDs in \cwdm pairs is still reasonable. For dMs, we infer masses from their spectral types \citep{Reid2005}. 

The resulting separations are projected linear separations and are upper limits for the true separations of these systems. The values of the binary separations should not be taken absolutely, but rather, as relative to each other to help group pairs into separation bins, i.e., very close ($<$0.1 au), close (0.1$-$1 au), wide (1$-$100 au). For more details on calculating and measuring the aforementioned parameters, see \cite{Morgan2012}.

We search the SDSS S82 footprint for matches to our initial sample of 1933 \cwdm pairs, and find 181 pairs that overlap within the S82 footprint. We obtain the light curves (based on the \cite{Ivezic2007} calibrations of Stripe 82 photometry) for each of these objects in {\it ugriz} filters. On average, each pair has more than 40 epochs. We calculate the quiescent magnitudes in each filter for the multi-epoch S82 dataset by converting the filter magnitudes into fluxes, finding the mean value in each filter for each star, performing a 3-sigma clip, and then recalculating a mean flux value in each filter after the sigma clipping. The sigma-clipped mean magnitude is then assigned as the quiescent magnitude. 

We apply similar quality cuts as prescribed in Section 2.1 of (K09), including: 1) requiring the quiescent flux of the star to be above the SDSS {\it ugriz} limiting magnitudes (22.0, 22.2, 22.2, 21.3, 20.5, respectively); and 2) removing objects where flags were set in the {\it u}- or {\it g}-band indicating bad photometry \cite[\texttt{SATURATED}, \texttt{NODEBLEND}, \texttt{NOPROFILE}, \texttt{PSFFLUX\textunderscore INTERP}, \texttt{BAD\textunderscore COUNTS\textunderscore ERROR}, \texttt{INTERP\textunderscore CENTER}, \texttt{DEBLEND\textunderscore NOPEAK}, \texttt{NOTCHECKED}; ][]{Stoughton2002b}. No stars were removed following these quality cuts, leaving a sample of 181 spectroscopic \cwdms with photometry for 9206 epochs from Stripe 82.

\subsection{Photometric Sample}
To increase our sample size of \cwdm pairs, and to ensure a statistical sample comparable to the field study (K09), which consists of both photometric and spectroscopic data from SDSS, we build a photometric sample of WD+dM. Previous studies have outlined photometric color cuts in the near ultra-violet \citep[{\it GALEX};][]{Martin2005}, optical (SDSS), and infrared \citep[e.g., 2MASS;][]{Skrutskie2006}, which can isolate close \cwdm pairs with $<20\%$ contamination \citep{Rebassa-Mansergas2014}. However, we first used a set of color cuts in {\it ugriz} to maximize the number of objects in our initial selection of photometric candidate \cwdm pairs. For the entire DR12 spectroscopic sample, the color cuts recovered 2369 objects, 1227 were spectroscopically confirmed to be \cwdm pairs (out of 1933 \cwdms known from previous visual inspection). The main contaminant in the \cwdm color-locus are broad-lined QSOs. If these objects are truly extragalactic sources, they should have no measurable proper motions. We require total proper motions $>$ 1.5 times the uncertainty in proper motion. This proper motion cut is chosen, somewhat arbitrarily, as we found it maximized the removal of the largest number of contaminants without removing too many of the spectroscopically confirmed \cwdm pairs. More stringent proper motion cuts were found to be provide only a marginal increase in effectiveness at removing contaminants while removing, fractionally, more spectroscopically confirmed WD+dM. We suspect this is in part due to the USNO-B proper motions not being as effective for faint objects residing in color-color regime occupied by \cwdm pairs, as shown in \cite{Theissen2015}. With the above proper motion quality cut and a photometric quality cut ({\it ugriz} uncertainties $<$ 0.1 mag), our photometrically selected candidate sample is reduced to 930 total objects and 765 spectroscopically confirmed \cwdm (out of 1166 previously known). Before the quality cuts, the color-selection recovers 63.4\% of known \cwdms with a 48.4\% contamination rate (objects other than \cwdm pairs); after the quality cuts they recover 65.6\% \cwdms with only a 17.8\% contamination rate.

Due to the specialized way that SDSS targeted objects spectroscopically \citep{Stoughton2002}, the distribution of objects in the spectroscopic sample may not be representative. As such, the above-quoted contamination rates for the color-cuts in the spectroscopic sample may not be perfectly reflected in the photometric sample. However, the spectroscopic selection algorithms employed by SDSS appear to preferentially select \cwdm pairs more than it would in a random selection process. It is likely that very few \cwdms remain in the photometric sample that have not already been targeted spectroscopically.

Nevertheless, we apply the color-cuts as well as the additional quality cuts as discussed above to the entirety SDSS Stripe 82 photometric sample. The color-cuts return 184 objects, 77 of which have spectroscopic observations. From the 77 objects that have spectroscopic observations, 55 are visually confirmed as \cwdms (32.5\% contamination), consistent with the findings in the spectroscopic sample. This leaves 105 objects selected by the color-cuts that don't have spectroscopic observations. As a secondary check, we visually inspect the {\it gri} composite images to check whether the photometric candidates were visually similar in color to spectroscopically confirmed \cwdm pairs. Only 30-35\% of the objects appear consistent with the spectroscopically confirmed {\it gri} composite images of \cwdms, the rest identified as either field dMs or extragalactic sources (i.e. broad-lined QSOs). Due to the low-fidelity and small sample size of the photometric sample, we choose not to include it in our analysis and concentrate on the spectroscopic sample where we are confident we are including bonafide \cwdm pairs.

\begin{figure*}[!htp]
   \centering
	  \includegraphics[trim=0.5cm 0.3cm 1cm 0.0cm,width=0.70\textwidth]{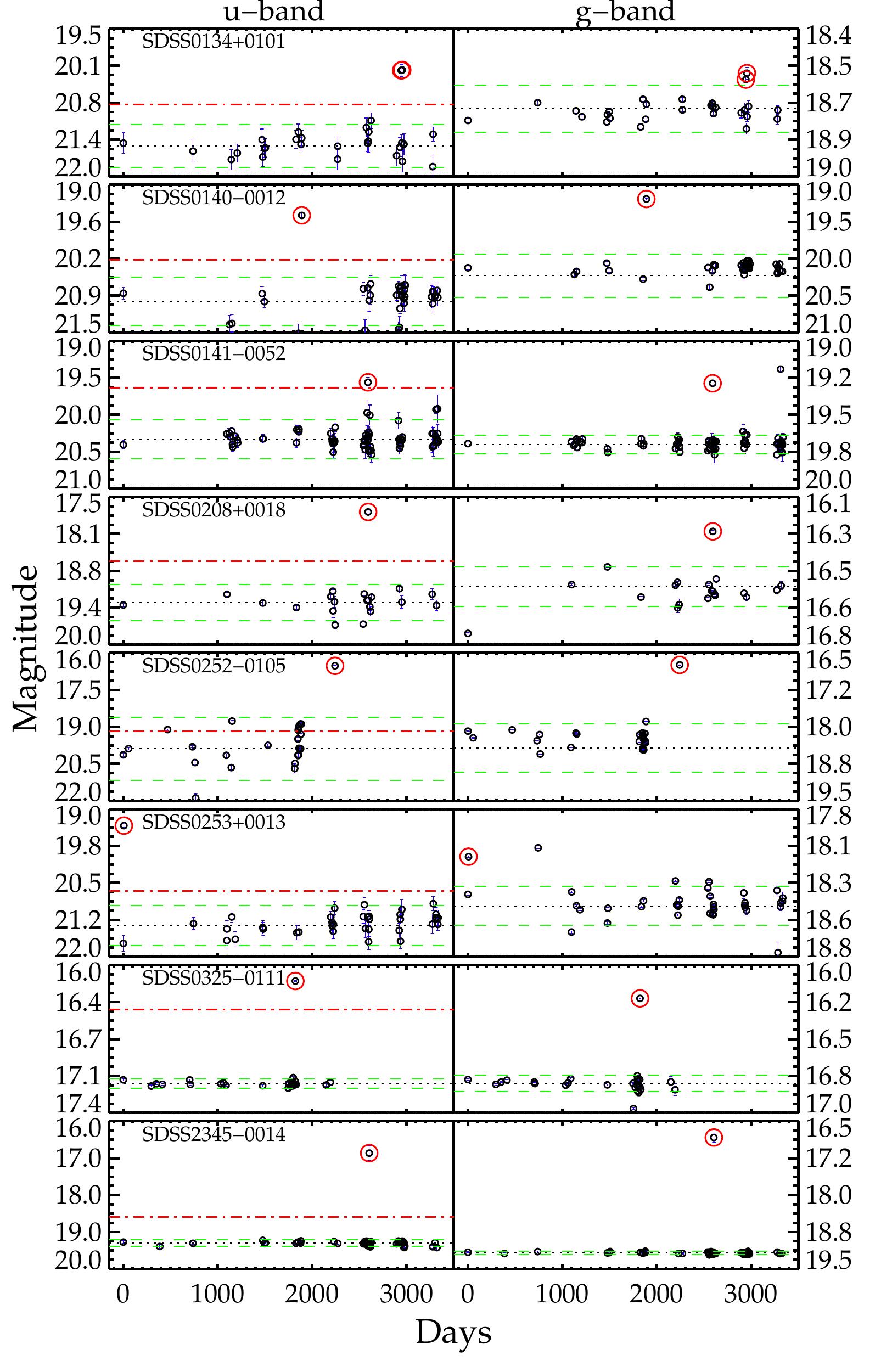}
   \caption{The light curves of all eight flaring close \cwdm pairs from the S82 survey. Each row is a separate object with the {\it u}-band in the left panel and the {\it g}-band in the right panel. Individual photometric measurements are plotted as black circles with their corresponding magnitude uncertainties plotted as blue error bars. The calculated quiescent magnitude is the black dotted line and the dashed green line above and below represents 2$\sigma$ in the quiescent magnitude. The red dash-dotted line in each of the left panels is the required $\Delta {\it u} > 0.7$ flux increase to rule out a false positive by the documented ``red-leak'' flaw in the {\it u}-band. Flaring epochs in both the {\it u}- and {\it g}-band are highlighted with red circles. To be classified as a flare, the following requirements must be met: 1) A flux increase 2$\sigma$ above the quiescent mean must be seen in both the {\it u}- and {\it g}-bands; 2) The flux increase must be greater than 0.7 magnitudes in the {\it u}-band; and 3) The candidate flaring epochs must be concurrent (separated by 108 seconds)} 
   \label{flare_ex}
\end{figure*}

\section{Flare Selection}\label{Sec:Flare Selection}
The purpose of this study is to compare the flare properties of field dMs and dMs with close binary companions. We use the K09 study as our comparison field dM flare sample, and follow their flare selection criteria as closely as possible. For our sample of 181 \cwdm pairs with light curves from Stripe 82, flares were identified by eye if they satisfied the following conditions:

\begin{itemize}
\item Each individual light curve must have $>$10 epochs in the {\it u}-band.
\item The {\it u}-band increase must be $\ge$ 0.7 magnitudes over the quiescent mean. This limit is imposed to reduce false positives resulting from the ``red leak,'' a well-known systematic effect in the SDSS {\it u}-band that can result in, at most a 0.5 magnitude increase (see Section~3.3 of Kowalski et al. 2009 for more information).
\item An increase in magnitude $>$2$\sigma$ above the quiescent mean must be seen concurrently in at least the {\it u}- and {\it g}-bands. For the {\it u}-band, this is in addition to the imposed $\Delta u \ge$ 0.7 requirement.
\item As before, we require that for all epochs the following standard photometric flags were not triggered: \texttt{SATURATED}, \texttt{NODEBLEND}, \texttt{NOPROFILE}, \texttt{PSFFLUX\textunderscore INTERP}, \texttt{BAD\textunderscore COUNTS\textunderscore ERROR}, \texttt{INTERP\textunderscore CENTER}, \texttt{DEBLEND\textunderscore NOPEAK}, \texttt{NOTCHECKED} \citep{Stoughton2002b}. Particularly important is the \texttt{NODEBLEND} flag, which is set when two objects were unable to be deblended and could masquerade as a flare.
\item For each flaring epoch, we download archival, raw {\it u}- and {\it g}-band images to look for any non-stellar related brightening (i.e., bad seeing, diffraction spikes from nearby bright stars, satellites, airplanes, etc.) that could masquerade as a flare.
\end{itemize}

We searched for flares on 181 stars in our \cwdm sample for which there were a total of 9206 concurrent epochs in both {\it u}- and {\it g}-bands (taken within 108 seconds of one another due to drift scan observing). In those 9206 epochs, we find nine flaring epochs across eight \cwdm systems (one \cwdm system flared twice). The flares in both the {\it u}- and {\it g}-bands for all eight flare stars are shown in Figure~\ref{flare_ex}. For each star, the {\it u}- and {\it g}-band light curves are shown side-by-side with each SDSS photometric measurement as a black circle and the corresponding photometric magnitude uncertainty as blue vertical error bars. In each panel, the black dotted line represents the calculated quiescent magnitude (as described in Section~2.2) and the green dashed lines represents a 2$\sigma$ deviation of the mean (excluding 2$\sigma$ clipped epochs). The flaring epochs are denoted by red circles and satisfy the criteria outlined above.

The top panel of Figure~\ref{flare_ex} shows the \cwdm SDSS0134+0101, where two flares are observed to be separated by $\sim$10 days. Typical stellar flares only last on the order of minutes to hours. As such, we treat these two events as independent flares. The notable flux increase (greater than 2$\sigma$ increase in both the {\it u}- and {\it g}-bands) in SDSS2057+0106 (third panel from the bottom) is not considered a flare because it narrowly misses the $>0.7$ magnitude increase requirement for the {\it u}-band. We observe similar events in other stars that narrowly miss at least one of our flare selection criteria. However, they were not included in our analysis to ensure that we could directly compare our sample to the K09 study.

\section{Sample Details}\label{Sec:Sample Details}

\begin{figure}[!ht]
   \begin{center}
      \includegraphics[trim=1cm 0.5cm 1.5cm 0.07cm,width=0.38\textwidth]{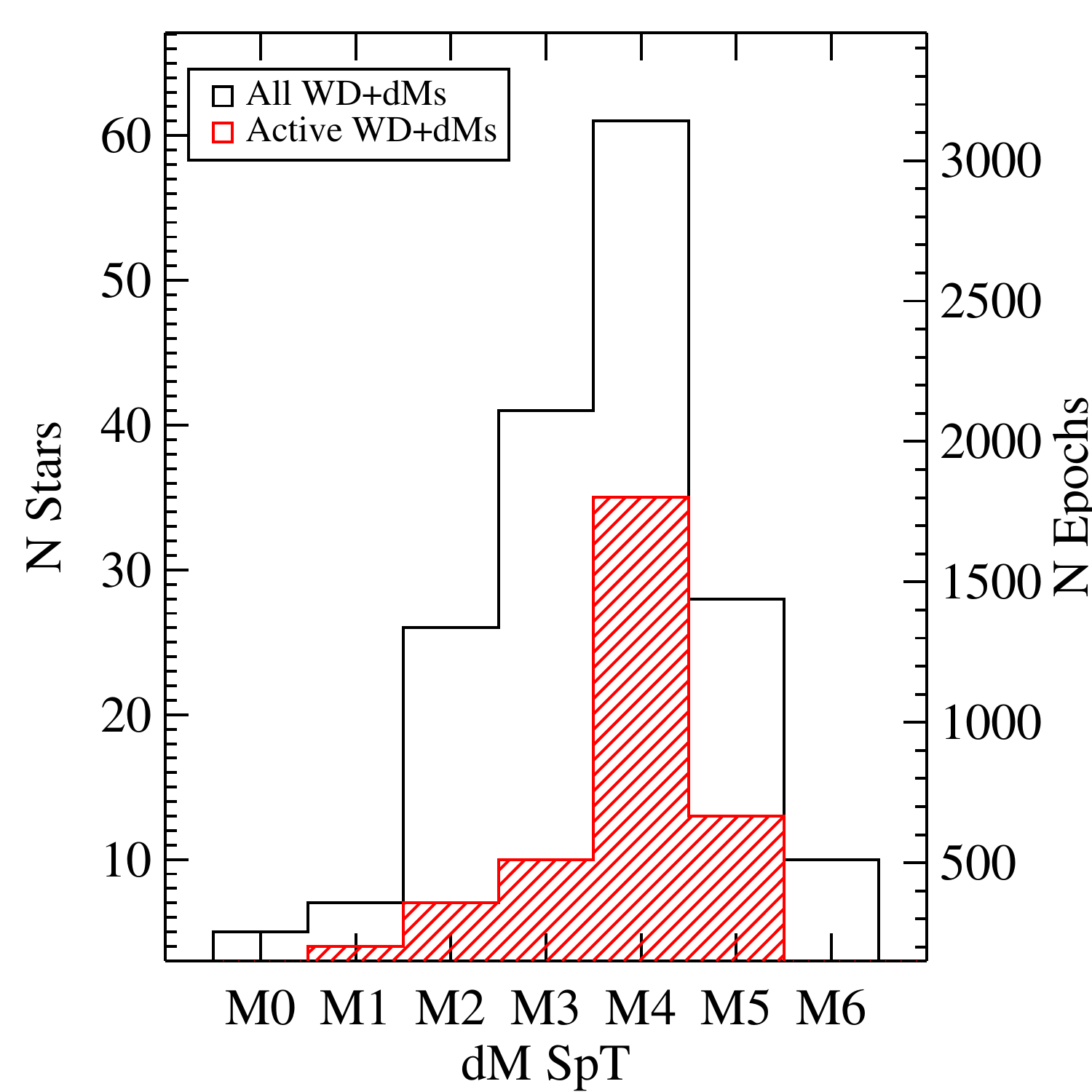}
      \end{center}
    \caption{The distribution of the total number of \cwdm pairs binned by spectral type (left y-axis) and the total number of epochs available from the S82 light curves binned by dM spectral type (right y-axis). The black line represents the entire sample, while the red dashed lined represents the epochs containing the active dMs (showing H$\alpha$ in emission).}
   \label{epochs}
\end{figure}

As mentioned above, our sample consists of 181 \cwdms, 71 of which are magnetically active as determined by the presence of H$\alpha$ emission in their spectra. There are 9206 total epochs around which we searched for flares, 3819 of which were from active stars. With a total of nine flares, two of which occur on active stars, we find flaring fractions of 0.09$\pm$0.03\% and 0.05$\pm$0.03\% for all \cwdms and active \cwdms, respectively. These flaring fractions are consistent with one another given the uncertainties (to less than one sigma).

The distribution of stars as well as the distribution of epochs as a function of spectral type are shown in Figure~\ref{epochs}. The black histogram is the \cwdm sample and the red dashed histogram denotes the active stars in the sample. Our sample distribution is consistent with what is expected from the dM initial mass function \citep[e.g.,][]{Bochanski2008, Bochanski2010}, where the distribution peaks around mid spectral types (M4-M5). One of the benefits of close \cwdm binary systems is that the two components have comparable luminosities, and different temperatures, and peak at different wavelengths; they are easily separated spectroscopically and sometimes photometrically. Unfortunately, this can lead to a selection effect in which early-type (M0-M1) and late-type ($\ge$ M7) dMs are under-represented. The higher contrast ratio between the more luminous early-type dMs and relatively cooler WDs ($\sim$10,000K) could lead to a dearth of early-type \cwdms. Conversely, the less luminous late-type dMs could be obscured by a WD with a relatively cool temperature ($\sim$10,000K).

\begin{figure}[!ht]
   \begin{center}
      \includegraphics[trim=1cm 0.5cm 1.5cm 0.0cm,width=0.40\textwidth]{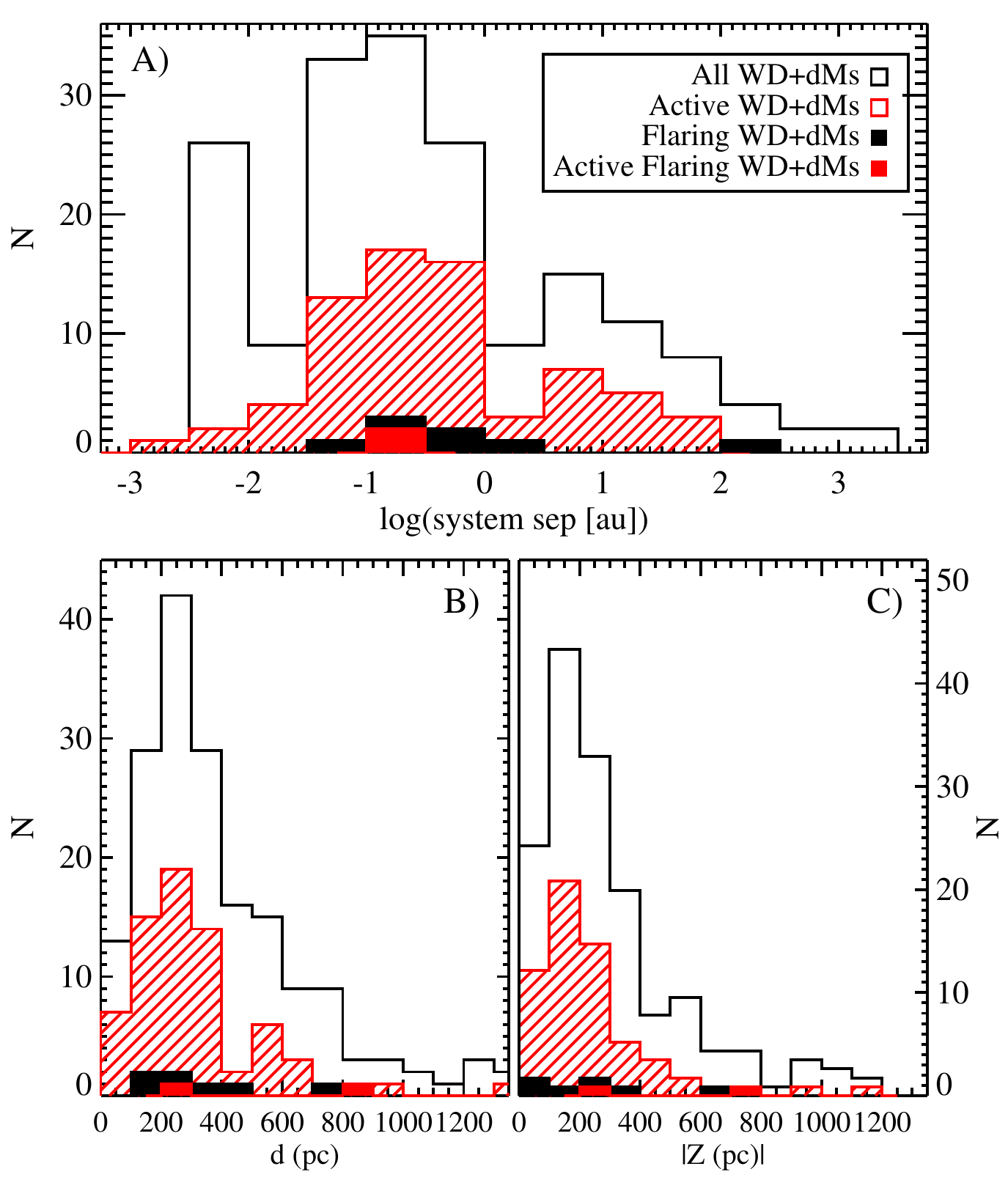}
      \end{center}
    \caption{Properties of the S82 \cwdm sample. Each panel is separated into all \cwdms (black open), active \cwdms (red dashed), all flaring \cwdms (black filled), active flaring \cwdms (red filled). A) Sample by projected physical separation [au], B) histogram of distance [pc], C) histogram of height above the Galactic Plane [pc].}
   \label{hists}
\end{figure}

Figure~\ref{hists} shows the distribution of projected physical system separations, absolute height above the Galactic plane, and distance to the S82 \cwdm sample. In every panel of Figure~\ref{hists} we show the entire \cwdm sample (black open), the active (as measured by H$\alpha$ emission) \cwdms (red thatched), the flaring \cwdms (black solid), and active flaring \cwdms (red solid). 

In the top panel of Figure~\ref{hists}, we show a distribution of the logarithmic projected physical system separation. In Section 2.1 we show that these are upper limits on the projected linear separations and that the values should not be used as absolute separations but rather used to compare relative separations (very close [$<$0.1 au], close [0.1$-$1au], wide [1$-$100 au]). The very close bin roughly corresponds to where tidal effects are thought to dominate \citep{Meibom2005}, whereas the middle and widest bins are thought to be where the binary companion is close enough to cause disks to be shorter-lived and thus, disrupt angular momentum loss mechanisms such as accretion powered stellar winds \citep[e.g.,][]{Matt2012} and magnetic disk locking \citep[e.g.,][]{Shu1994}.

The lower left panel of Figure~\ref{hists} shows distances calculated using spectroscopic parallax relations derived from \cite{Bochanski2008}. While photometric parallax relations are more reliable for dMs, the flux contamination from the WD makes these relations unreliable for our dMs. We estimate the spectroscopic parallax relations have $\sim$20\% precision \citep{Bochanski2011}. In the lower right panel, we present the absolute height above the Galactic plane, which has been shown to be a proxy for stellar age \citep{West2004, West2008, West2011}. In general, stars that orbit farther from the Galactic plane are older, as they have had more time to be dynamically heated. The entire \cwdm S82 sample (black open line), as well as the eight flaring stars (solid black lines), show a relatively even spread in distance and the absolute Galactic height above the plane, indicating no obvious age biases. 

Table~1 lists the flaring \cwdms. We list the SDSS identifier, dM spectral type, H$\alpha$  EW, activity state, physical projected separation, distance, absolute height above the galactic plane, the change in the {\it u}-band magnitude during a flare ($\Delta${\it u}$_{\textrm{flare}}$), and the flare luminosity in the {\it u}-band (L$_{\textrm{u,Flare}}$). Using the spectroscopic parallax distances, we estimate L$_{\textrm{u,Flare}}$ from the luminosity of the \cwdm system in quiescence subtracted from the luminosity of the \cwdm system in a flaring state. For the quiescent luminosity we use the quiescent, sigma-clipped mean from the light curve as discussed in Section~3.

\begin{deluxetable*}{lcrrrrrrr}
   \centering
   \tablewidth{0pt}
   \tablecolumns{9}
   \tabletypesize{\scriptsize}
   \tablecaption{Flaring Sample}
   \tablehead{\colhead{Object ID} &
   		    \colhead{dM SpT} &
		    \colhead{H$\alpha$ EW\tablenotemark{1}} &
		    \colhead{Active} &
		    \colhead{s} &
		    \colhead{d} &
		    \colhead{$|$Z$|$} &
		    \colhead{$\Delta${\it u}$_{\textrm{flare}}$} &
		    \colhead{log(L$_{\textrm{{\it u},flare}}$)} \\
		    \colhead{} &
		    \colhead{} &
		    \colhead{} &
		    \colhead{} &
		    \colhead{(au)\tablenotemark{2}} &
		    \colhead{(pc)\tablenotemark{3}} &
		    \colhead{(pc)\tablenotemark{4}} &
		    \colhead{} &
		    \colhead{(erg s$^{-1}$)\tablenotemark{5}} }
\startdata
        SDSS0134$+$0101 & 0.5 & \nodata & no &   0.17 &  719 &  607 &  1.28 & 30.23 \\
        & & & & & & &  1.28 & 30.24 \\
        SDSS0140$-$0012 & 2.5 &  3.15 & yes &   0.16 &  847 &  723 &  1.45 & 30.68 \\
        SDSS0141$-$0052 & 4.5 &  4.03 & yes &   0.17 &  270 &  221 &  0.77 & 29.51 \\
        SDSS0208$+$0018 & 2.5 &  0.20 & no &   0.98 &  115 &   81 &  1.53 & 29.66 \\
        SDSS0252$-$0105 & 4.0 & \nodata & no &   0.42 &  119 &   77 &  3.37 & 30.28 \\
        SDSS0253$+$0013 & 1.5 & \nodata & no & 177.71 &  398 &  289 &  2.03 & 30.15 \\
        SDSS0325$-$0111 & 3.5 & \nodata & no &   0.05 &  284 &  186 &  0.97 & 30.98 \\
        SDSS2345$-$0014 & 4.0 &  0.63 & no &   1.41 &  401 &  328 &  2.43 & 31.17 \\
\enddata
\tablenotetext{1}{For missing data, no measurement of H$\alpha$ EW could be made.}
\tablenotetext{2}{The separations were calculated using the dM and WD RVs after being corrected for system RVs. We assumed that the measured absolute RVs of each component were the orbital velocities as well as assuming edge-on, circular Keplarian orbits. See Section~4 for more details.}
\tablenotetext{3}{Distances were calculated using a spectroscopic parallax relation derived from data found in \cite{Bochanski2008}. The expected uncertainty in these calculations is 20\%}
\tablenotetext{4}{Galactic height above the plane is calculated using Eq. 9 in \cite{Morgan2012}}
\tablenotetext{5}{Flare luminosities are calculated by subtracting the quiescent luminosity from  luminosity of the system while in a flaring state. We used the distances found in this table to estimate luminosities.}
\end{deluxetable*}
\label{tab_flaresamp}

We consider the possibility of an interloping foreground or background dM, undergoing a flare, that could be contaminating our flaring close \cwdm sample. We investigate this possibility by empirically calculating the chance for a foreground or background dM to be within 1.5$^{\prime\prime}$ of each of our eight flaring stars. This is done by counting the number of objects from the SDSS DR12 photometric database that have similar colors to dMs, within a 1x1 square degree field of view centered on each of our targets. For the colors, we require {\it r} $-$ {\it i} $\ge$ 0.3 and {\it i} $-$ {\it z} $\ge$ 0.2,  with {\it r} $<$ 22, {\it i} $<$ 22, {\it z} $<$ 21.2, as well as requiring {\it r} $>$ 16 to rule out any M giants. These color cuts include objects that are slightly bluer than most dMs and are likely to include a small portion of galaxies, leading to overestimated stellar counts. Given the stellar counts in each 1x1 square degree field, we calculate the number of objects to fall within 1.5$^{\prime\prime}$ of our target using the following equation: number = $\pi \times (1.5^{\prime\prime})^2 \times \textrm{counts}/\textrm{degree}^2 * (1~\textrm{degree}/ 3600^{\prime\prime})^2$. For each of our flaring WD+dM pairs, the highest probability that there would be a foreground dM within 1.5" was 0.5\% (with the mean probability being  $\sim$0.1\%). Thus, we do not expect any of our flaring \cwdm pairs to be due to a foreground flaring dM.

One physical process that could affect the statistics of this sample of \cwdms is any potential mass transfer onto the dM during the common-envelope phase, thereby altering the dM natal mass and adding uncertainty to our spectral type classifications. Not much is known about common-envelope phase as the time-scale is very short (on the order of tens of years), making this stellar evolution process difficult to observe in order to place empirical constraints \citep[e.g.,][]{Ivanova2013}. However, models have shown that while the companion star is engulfed by the common-envelope, unless it is a degenerate star, it is unlikely to accrue any mass from the surrounding envelope \citep{Ivanova2013, Webbink1988, Hjellming1991}. Thus, we expect our spectral type classifications to be representative of the dM mass. In addition, we do not see any evidence that these binaries are near-contact and experiencing ongoing mass-accretion. Only two stars show H$\alpha$ emission and there is no evidence for any broadening due to mass-accretion (see Figure~\ref{ha_flares}).

In addition, one potential source of selection bias in our sample is the flux boosting provided by the WD, specifically in the $u$-band. Without the presence of the companion WD, the dM may never have been included in the sample ($u < 22.0$). Thus, we are effectively probing a larger volume than in the K09 study. This, however, should bias our sample towards older, and more inactive systems, which are less likely to flare (given that we are looking out of the Southern Galactic plane). If flare rates correlate with age (e.g., K09), as does magnetic activity \citep{West2008}, then our derived \cwdm flare rate would be underestimated. Our systems show no obvious age bias in distance or height above the Galactic plane (panels b and c in Figure~\ref{hists}), so we expect this effect to be minimal.

\section{Results and Discussion}\label{Sec:Results}
With only eight \cwdm flaring systems (nine total flares), a robust comparison between the field and binary populations is difficult. However, this small sample can still provide insights on how close binarity affects dM flaring behavior. 

\begin{figure*}[!ht]
   \begin{center}
      \includegraphics[trim=0.0cm 0.5cm 0.0cm 0.0cm,width=0.80\textwidth]{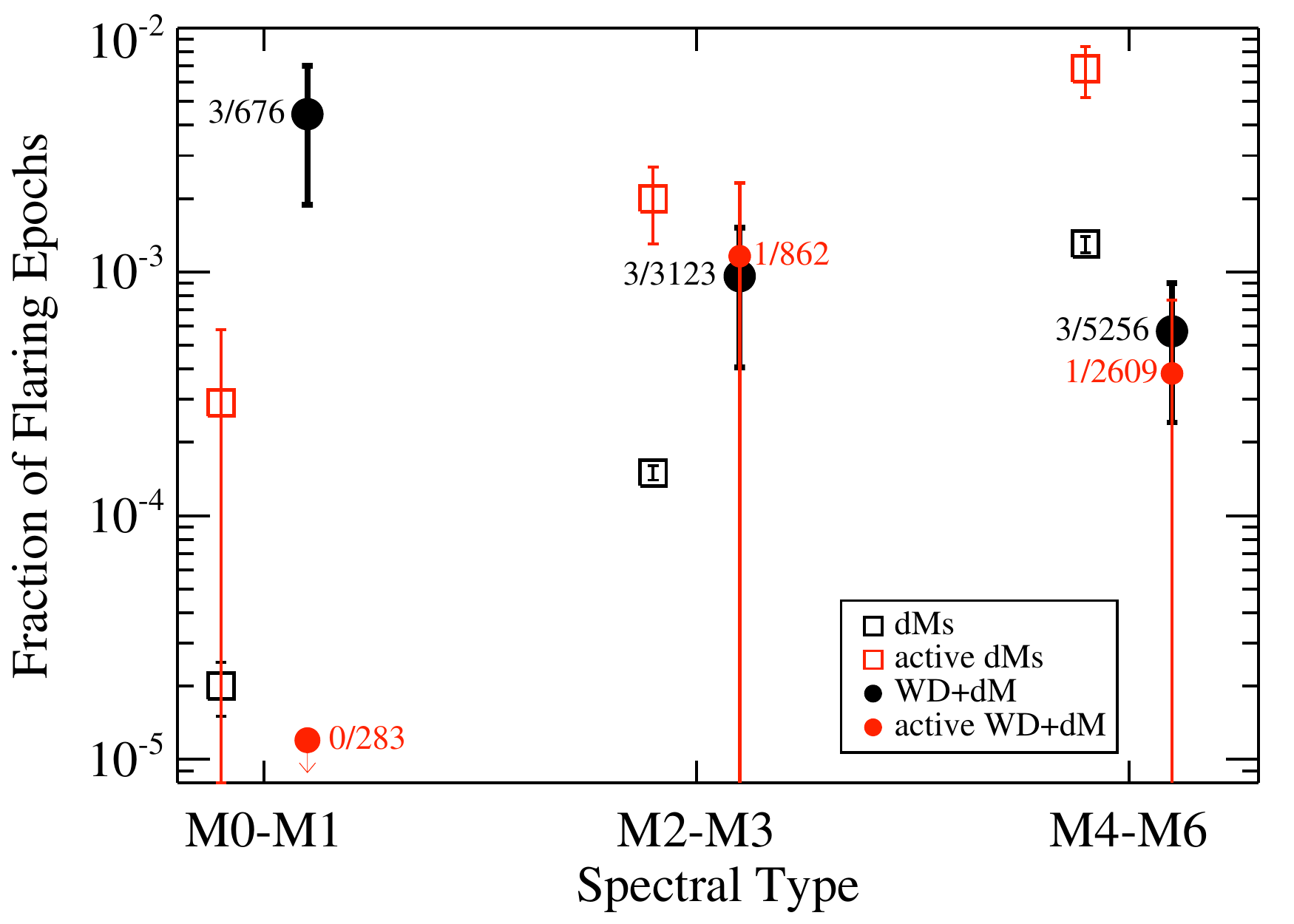}
      \end{center}
    \caption{Fraction of flaring epochs as a function of binned dM spectral type. Field dMs (open black squares), active field dMs (red open squares), all \cwdm pairs (filled black circles), active \cwdm pairs (filled red circles); error bars are calculated using the binomial distribution. For early (M0-M1) and mid-type (M2-M3) spectral types, the \cwdm pairs flare more frequently than the field dMs, while late-type \cwdms flare at a similar rate (to within the errors) than the field dMs. Only two of the eight \cwdms are observed to be active, one in the mid-type spectral type bin and one in the late spectral type bin. We hesitate to make a comparison between the active dM and \cwdm flaring fraction due to having only one active \cwdm in each of the M2-M3 and M4-M6 spectral type bins.}
   \label{flareplot}
\end{figure*}

Figure~\ref{flareplot} shows the fraction of flaring epochs as a function of dM spectral type for field dMs (open black squares), active field dMs (open red squares), \cwdms (black filled circles), and active \cwdms (red filled circles). The data presented for field dMs is from Figure 9 of K09. In Figure~\ref{flareplot}, we see a trend of increasing flaring fraction for the field dMs as a function of later spectral types, as we would expect given prior dM activity studies \citep[e.g.,][]{West2011}; late-type dMs are more likely to be active than early-type dMs. In addition, the presence of H$\alpha$ emission, or activity, is well known to correlate with flares on dMs. Then, it is not surprising that active dMs show a higher flaring fraction than the full dM sample (both active and inactive). 

Early-type \cwdms (M0-M1) flare 220 times more frequently than the field dM population and the active \cwdms flare 15 times more frequently than the active field population. These increases in flaring fractions are statistically significant at the 1.5-sigma level. At the mid spectral type bin (M2-M3), the \cwdm pairs flare 6.5 times more frequently than the field dM sample, significant again at 1.5-sigma level. While the active \cwdms and the active field dMs flare at approximately the same frequency, to within the binomial errors. For the late-type stars (M4-M6), we see a different trend, the \cwdms are flaring 50\% as frequently as the entire dM sample and only 8\% as frequently as the active field dMs, at the 2-sigma and 3.8-sigma levels, respectively. 

\begin{figure}[!ht]
   \begin{center}
      \includegraphics[trim=1cm 0.5cm 1cm 0cm,width=0.45\textwidth]{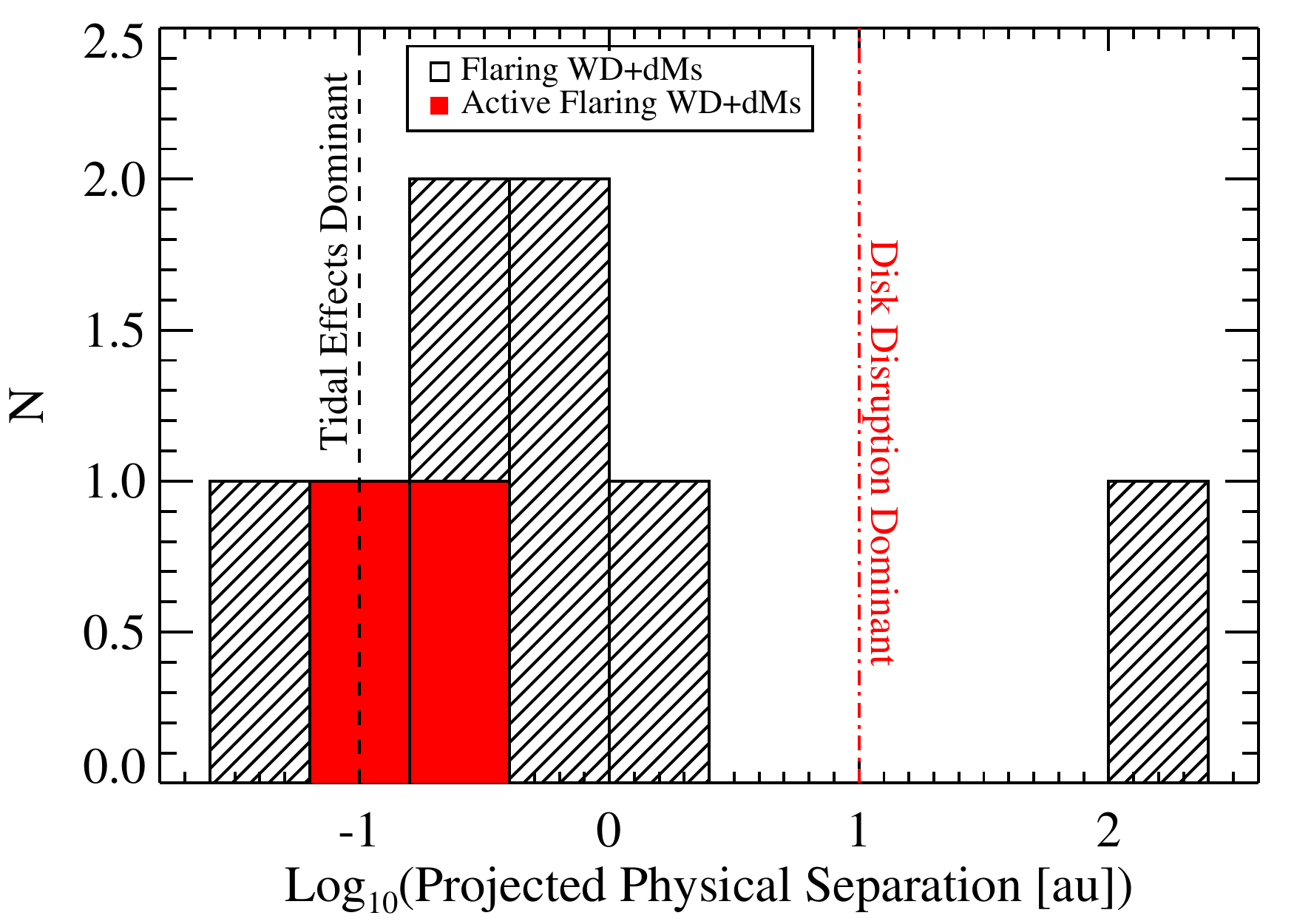}
      \end{center}
    \caption{Distribution of the flaring \cwdm sample and the estimated projected physical separations (au). The black lined histograms show all of the \cwdm pairs while the red solid histograms highlight the two active pairs. The dashed vertical line at 0.1 au is where tidal effects are thought to dominate while the dash-dotted line at 10 au is where disk disruption effects are thought to be dominant.}
   \label{sep_plot}
\end{figure}

Our results indicate that close companions likely do affect the flaring rates of early-type dMs. Similarly, in \cite{Morgan2012}, the \cwdm activity fractions were enhanced above the field dM population at early spectral types and then decreased to comparable activity fractions in later spectral-types. The activity enhancement seen in the \cwdm pairs was attributed to the close binary quenching angular momentum loss-mechanisms, primarily through shortened disk lifetimes, causing the dMs to rotate faster than field dMs of similar ages (and masses). This is corroborated in our current study by Figure~\ref{sep_plot}, which shows a histogram of the system separations of the flaring \cwdm pairs. In general, all of the pairs are relatively close ($<$1 au). Some of these systems are in a regime where tidal effects likely play a role in impeding stellar spin down and are rotating faster than field dwarfs of similar ages and masses. Interestingly, the only two active systems also have two of the closest separations. As mentioned in Section~4, the exact binary separations have large uncertainties and should only be taken as rough relative guides for the binary separations.

\begin{figure}[!ht]
   \begin{center}
      \includegraphics[trim=0.5cm 0.5cm 1cm 0cm,width=0.50\textwidth]{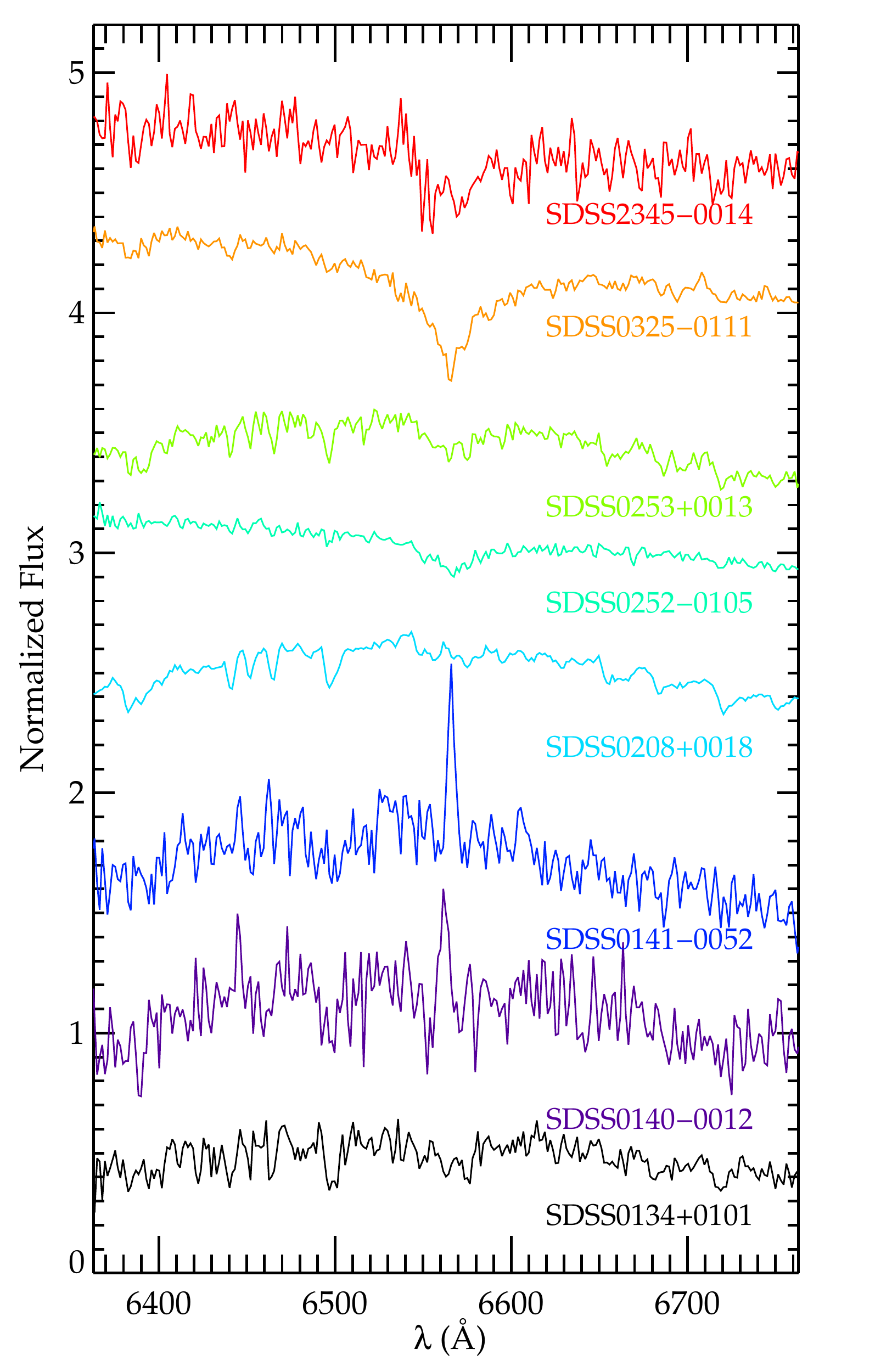}
      \end{center}
    \caption{Normalized flux of the SDSS spectra for all 8 close \cwdm flaring stars centered on the H$\alpha$ wavelength region. The activity state was determined by-eye and only SDSS0140-0012 and SDSS0141-0052 are reported as active (as measured by H$\alpha$ emission). The spectra displayed are the raw spectra and have not had the white dwarf component subtracted.}
   \label{ha_flares}
\end{figure}

We are not confident in making any spectral-type-activity dependent statistical comparisons to the K09 flaring sample because only two of our eight \cwdm flaring stars are active (H$\alpha$ is present in emission). However, only 8\% of the field dM flaring sample showed evidence for inactivity (no H$\alpha$ in emission) while we report 75$^{+9}_{-19}$\% of our sample as inactive (shown in Figure~\ref{ha_flares}). Magnetically inactive stars have been shown to flare in the past, albeit at lower rates than active stars. Our sample of mostly inactive flaring systems is puzzling. 

We perform a simple statistical test to determine whether or not this trend could be due to the small number statistics associated with our sample. Assuming that 8\% of all flaring dMs are inactive, as reported by K09, we use a binomial probability distribution to determine the likelihood that we pull six inactive stars out of eight draws and find the probability of this outcome to be 2.3\%. Thus, with 97.7\% certainty ($>$2$\sigma$), we do not expect our reported inactivity fraction of 75$^{+9}_{-19}$\% to be due to small number statistics.

\subsection{Effect on the Galactic transient background}
Quantifying dM flare rates is important for transient Galactic and extragalactic studies because it places empirical constraints on the noise in the Galactic transient background, which is mostly due to stellar flares \citep[e.g.,][]{Becker2004, Rykoff2005, Kulkarni2006, Rau2008, Berger2013}. The dM flare rate will also change depending on the Galactic sight line; closer to the Galactic plane will have a higher flare rate than will farther above the Galactic plane (K09). Precise metrics of flare rates around dMs, such as FFDs, require costly short-cadence and long-time observational programs. We use the coarse time-series data from the SDSS Stripe 82 to estimate Galactic flare rates for close \cwdm pairs as was done for field dMs in K09. Due to the sparseness of the SDSS Stripe 82 data, we have no information on the duration of the flares. Therefore, our following discussions of flare rates refer to the likelihood of seeing the stars in a flaring state, rather than the number of flares observed over time. As such, our flare rates are analogous to those reported in \cite{Kowalski2009} in which they are estimates for the lower-limit of the flaring rate.

In the S82 field dM study, the lower-limit of the Galactic flare rate was estimated to be 1.3 flares hr$^{-1}$ deg$^{-2}$ (for flares $\Delta${\it u} $\ge$ 0.7 and {\it u} $<$ 22). The estimate was made by taking the total number of dM flares found (271; K09), dividing the observing time, $\sim$50 epochs on average per star (54.1s per exposure), and dividing by the area sampled by the S82 survey (271.9 deg$^{2}$). The reported field dM flare rate is averaged over the entire S82 footprint and will likely vary depending on the Galactic sight line; K09 estimate rates as high as 8 flares hr$^{-1}$ deg$^{-2}$ for the most crowded low-latitude sight lines.

Using a similar procedure to K09, we calculate an estimate for the lower-limit of flare rates for the close \cwdms in our sample. Using our 9 flares, with an average of 50 exposures per star, and 271.9 deg$^{2}$, we report an estimated lower-limit flaring rate of 0.04 flares hr$^{-1}$ deg$^{-2}$ (for flares $\Delta${\it u} $\ge$ 0.7 and {\it u} $<$ 22). This results in only a 3\% contribution to the estimated flare rate of field dMs. Our results show that individual, early-to-mid spectral type \cwdm pairs are more likely to flare than field dMs, however, they are simply not as numerous as field dMs to significantly contribute to total number of Galactic dM stellar flares. We conclude from this study that the \cwdm flare rates are a negligible contribution to the Galactic transient background in relation to field dMs. However, \cwdms can be used as a proxy for close binaries, in general. 

We use \cwdms in this study, as well as in \cite{Morgan2012}, to build a large statistical sample of dMs with close, unresolved binary companions. We expect that most non-WD close companions to dMs will cause similar activity enhancements, as these seen with close WD companions \citep{Morgan2012}. Realistically, the activity enhancement will likely depend strongly on the separation of the system and the mass of the companion to the dM, a relationship that is largely unconstrained and difficult to determine empirically. Recent studies show possible increases in dM X-ray activity (a proxy for magnetic activity) due to a closely orbiting giant planet \citep[e.g.,][]{Poppenhaeger2011, Poppenhaeger2011a, Poppenhaeger2014}, so we suspect that this activity enhancement can extend to very lower-mass companions (i.e. the planet regime). From new high-resolution optical studies using adaptive optics imaging (AO), \cite{Ward-Duong2015} find that 21$\pm$3\% of dMs have a close, low-mass companion ($\le$ 0.30 M$_{\odot}$) within a 1$-$100 au projected physical separation. This is corroborated by prior studies that found similar multiplicity fractions for close binary separations among dMs \citep[e.g.,][]{Fischer1992, Janson2012, Janson2014}. The previously quoted multiplicity fraction is likely an underestimate due the studies being insensitive to extremely close binary systems ($<$1 au) and ignoring dMs with higher-mass main sequence companions (progenitors to \cwdm systems). 

If we assume that $\sim$20\% of dMs have a low-to-equal mass close binary companion \citep[e.g.][]{Fischer1992, Janson2012, Janson2014, Ward-Duong2015}, most of which go unnoticed and cause enhancements in magnetic activity, then, multiplicity has important implications for the dM flare rate. Additionally, the field dM flare rate reported by K09 may actually be the total dM {\it system} flare rate. That is to say, the flare rate reported by K09 is the single + close binary dM flare rate since identifying close binaries from SDSS photometry is extremely difficult. As such, using our measured flare rate of \cwdms as a proxy for the close binary flare rate, we can perform a quick calculation to estimate the single dM flare rate:
\begin{eqnarray}
\begin{array}{ll}
FR_{total,norm} \times N_{total} =  \\
FR_{single,norm} \times N_{single} + FR_{binary,norm} \times N_{binary},
\end{array}
\end{eqnarray}
where:
\begin{itemize} \itemsep0.2pt \parskip0pt \parsep0pt
	\item[--] FR$_{total,norm}$ = (2.7 $\pm$ 0.2)$\times$10$^{-5}$ flares deg$^{-2}$ hr$^{-1}$ system$^{-1}$. This is the total flare rate that includes both field dMs and dMs with unseen close binary companions (not only just WD companions). FR$_{total,norm}$ is derived from FR$_{total}$ = 1.3 flares deg$^{-2}$ hr$^{-1}$ (reported by K09), normalized by the number of stars searched over in the K09 study (50130). We use poisson statistics to estimate the uncertainties in the normalized flare rate.
	\item[--] FR$_{binary,norm}$ = (2.4 $\pm$ 0.8)$\times$10$^{-4}$ flares deg$^{-2}$ hr$^{-1}$ system$^{-1}$. This value is derived from FR$_{\cwdm}$ = 0.04 flares deg$^{-2}$ hr$^{-1}$, which is the flare rate for \cwdms. By normalizing FR$_{\cwdm}$ by the total number of \cwdm searched over in our study (181), this value now becomes FR$_{binary,norm}$, an estimate for the flare rate for all dMs with close binary companions, not only dMs with close WD companions. Again, we use poisson statistics to estimate the uncertainties.
	\item[--] FR$_{single,norm}$ is the single field dM flare rate.
        \item[--] N$_{total}$ is the total number of systems including field dMs and dMs with close companions.
        \item[--] N$_{binary}$ is the number of dMs with an unseen, close binary companion. Following \cite{Ward-Duong2015}, we assume N$_{binary}$ = (0.2 $\pm$ 0.03)$\times$ N$_{total}$. 
        \item[--] N$_{single}$ is the number of single, field dMs. We set N$_{single}$ = (0.8 $\pm$ 0.03) $\times$ N$_{total}$, adopting the same 3\% uncertainty as the binary fraction. We assume, as mentioned above that 80\% of the dMs found in SDSS Stripe 82 are field dMs, while the other 20$\pm$3\% have unseen, close low-mass companions as predicted by \cite{Ward-Duong2015}.
\end{itemize}
Using the above normalized flare rates we calculate a FR$_{single,norm}$ = ($-$2.7 $\pm$ 2.0)$\times$10$^{-5}$ flares deg$^{-2}$ hr$^{-1}$ star$^{-1}$, which is consistent with a flare rate of zero (at the 1.35-sigma level).

Unexpectedly, our simple estimation returns a negative flare rate for single dMs. At face value, our results indicate that flare rates around dMs with close binary companions dominate the field dM flare rates. A FR$_{single,norm}$ $\le$ 0 is not realistic, as we know from numerous studies that isolated, young, active dMs flare often \citep[e.g.,][]{Lacy1976, Hilton2011, Hawley2014}, while older, inactive dMs flare much less often \citep{Hilton2011, Hawley2014}. Both K09 and this study utilize SDSS S82, which look away from the plane of the Galaxy towards the Southern Galactic Cap and will be biased towards older, most distant systems. We know from other studies that magnetic activity is short-lived in early-type dMs \citep[$\sim$1 Gyr for M0-M2 and $\sim$2$-$4 Gyr for M3-M4;][]{West2008} and that the K09 study could contain largely inactive and older, early-type dMs ($<$M3). In this scenario, the flares being reported in the early-type dMs from K09, and other similar studies \cite{Hilton2011}, may be dominated by the early-type dMs with unseen close companions. This may be one explanation for why we conclude that close binaries dominate the total dM flare rate reported by K09.

Unfortunately, our above analysis depends on the following poorly constrained parameters: 1) the close binarity fraction of dMs (limited to separations $>$1 au), 2) the efficiency of activity enhancement in dMs due to a close companion (assumed here as 100\%); and 3) the close binary flare rate. To return to a realistic, positive single dM flare rate, we can use the results presented here to constrain the upper limits of the close binarity fraction (N$_{binary}$) and the close binary flare rate (FR$_{binary,norm}$). To do this, we ask the following questions: 1) what fraction of binary stars (N$_{binary}$) do we need to calculate a single dM flare rate (FR$_{single,norm}$), if we assume FR$_{binary,norm}$ is true? Similarly, 2) what binary star flare rate (FR$_{binary,norm}$) do we need to calculate a positive single dM flare rate (FR$_{single,norm}$), if we assume N$_{binary}$ is true?  

For the first question, we solve Equation 1 for N$_{binary}$ by setting FR$_{single,norm} \ge$ 0, FR$_{binary,norm}$ = (2.4 $\pm$ 0.8)$\times$10$^{-4}$ flares deg$^{-2}$ hr$^{-1}$ system$^{-1}$, and FR$_{total,norm}$ = (2.7 $\pm$ 0.2)$\times$10$^{-5}$ flares deg$^{-2}$ hr$^{-1}$ system$^{-1}$. We calculate N$_{binary} \le$ (0.11 $\pm$ 0.03). For the second question, we solve Equation 1 for FR$_{binary,norm}$ by setting FR$_{single,norm} \ge$ 0, N$_{binary}$ = (0.20 $\pm$ 0.03) $\times$ N$_{total}$, FR$_{total,norm}$ = (2.7 $\pm$ 0.2)$\times$10$^{-5}$ flares deg$^{-2}$ hr$^{-1}$ system$^{-1}$; we calculate FR$_{binary,norm}$ $\le$ (1.3 $\pm$ 0.2)$\times$10$^{-4}$ flares deg$^{-2}$ hr$^{-1}$ system$^{-1}$. This corresponds to only a 54\% reduction in either parameter, which is reasonable due to N$_{binary}$ being an upper-limit and assuming 100\% efficiency of a close companion within 1$-$100 au in enhancing dM magnetic activity and flare rates. This efficiency is still poorly constrained and is likely much less than 100\% as it depends on both the mass of the companion and separation. Similarly, a 54\% reduction in FR$_{binary,norm}$ is within reason considering our total \cwdm flaring fractions have a 33\% uncertainty (0.09$\pm$0.03\%).

\subsection{Effect on the Habitability of Alien Worlds}
Since the launching of the {\it Kepler} spacecraft, almost two thousand exoplanets have been confirmed, with thousands more planet candidates identified (exoplanets.eu; NASA Exoplanet Archive). Despite this large sample, only a handful of exoplanets have been found in relatively close binary systems, most notably the circumbinary (orbiting both stars) planets {\it Kepler}-16b \citep{Doyle2011}, {\it Kepler}-34 \& 35 \citep{Welsh2012}, {\it Kepler}-38b \citep{Orosz2012a}, and {\it Kepler}-47 \cite[hosting two exoplanets,][]{Orosz2012}. What factors control the true habitability of exoplanets around other stars is still a topic of ongoing research. Even less is known about the potential for habitable planets in binary star systems, circumbinary or otherwise. Our study informs one aspect of habitability, namely, irradiation of exoplanet atmospheres by energetic events such as flares. Our data suggest that, at the very least, early-type (M0-M1) and mid-type (M2-M3) dMs in close binary systems flare $\sim$220 and $\sim$6 times more frequently, respectively, than their their field dM counterparts.

\begin{figure}[!ht]
   \begin{center}
      \includegraphics[trim=1cm 1cm 0cm 1cm,width=0.48\textwidth]{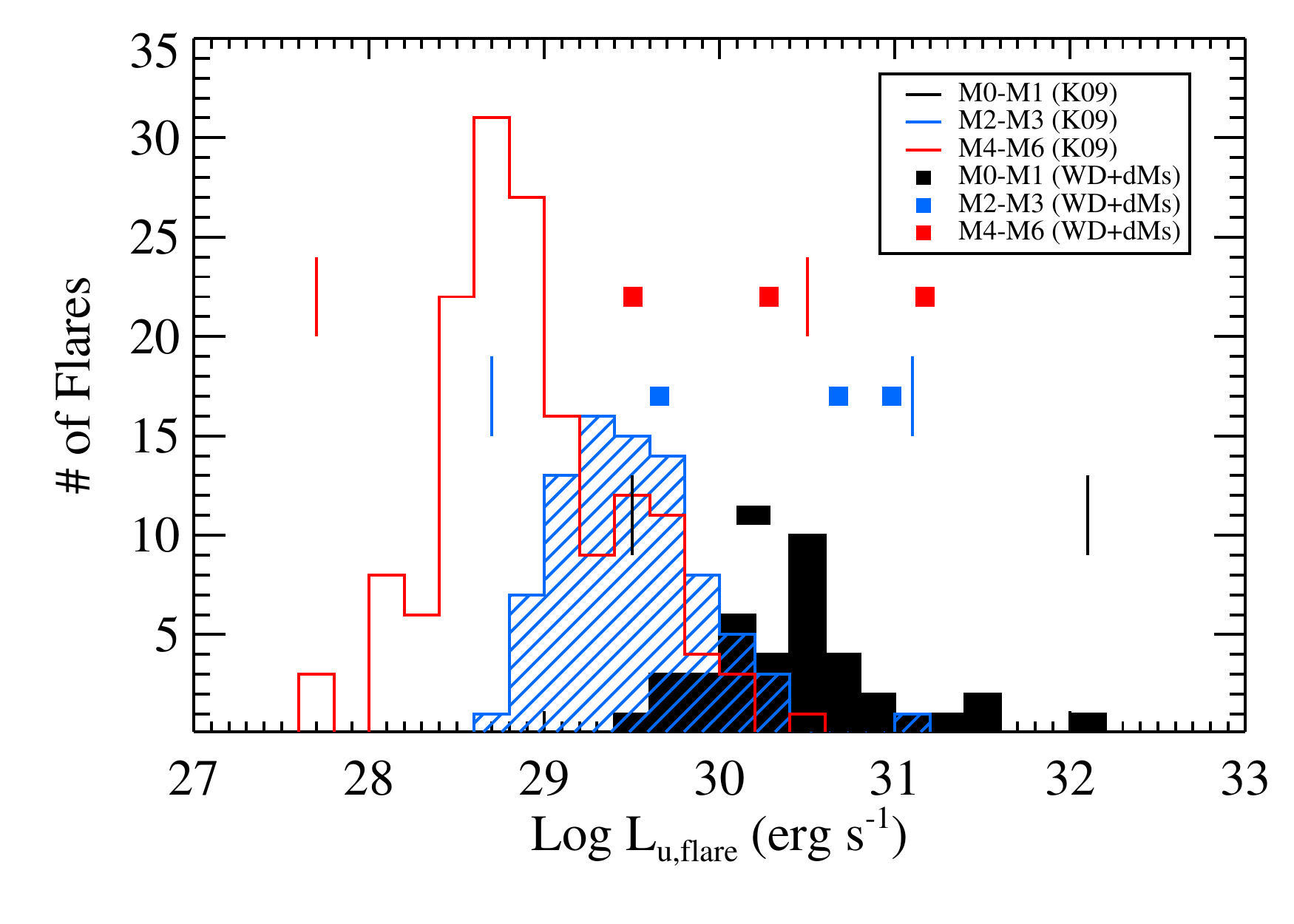}
      \end{center}
    \caption{Distribution of log(L$_{u,\textrm{flare}}$) [erg s$^{-1}$] as measured by the {\it u}-band magnitude increases binned by spectral type. The histograms are the flare luminosities for the field dM population taken from Figure 11 of K09. The flare luminosities are grouped into three spectral type bins early-type (M0-M1, filled black histogram), mid-type (M2-M3, hatched blue histogram), and late-type (M4-M6, open red histogram). The colored vertical lines bracket the log(L$_{u,\textrm{flare}}$) for each of the spectral type bins. The \cwdm flare luminosities are binned by spectral type in a similar fashion, early-type (M0-M1, red filled square), mid-type (M2-M3, blue filled square), late-type (M4-M6, black filled square). With the exception of one flare in a late-type \cwdm system, all of the measured flare luminosities fall within the bounds of the field dM sample. There is some evidence for systematically higher energies in early- and mid-type \cwdms, but, from these data alone we do not that conclude \cwdms on average produce higher luminosity flares.}
   \label{lumflares}
\end{figure}

In addition to the flaring frequency, another important factor in understanding how exoplanet atmospheres might be affected by stellar flares is the energy that is released during a flare event. We list the estimated luminosities (log(L$_{u,\textrm{flare}}$) [erg s$^{-1}$]) of the flares in our sample in Table~1. We compare our log(L$_{u,\textrm{flare}}$) values to these of (K09; Figure 11) in Figure~\ref{lumflares}. Flare luminosities are binned by early (M0-M1, black), mid (M2-M3, blue), and late (M4-M6, red) spectral types. The field dM sample is shown in histograms, while the \cwdm sample is shown as filled squares. We add colored vertical lines (for each spectral type) bracketing the total range of log(L$_{u,\textrm{flare}}$) of the field dM sample.

From Figure~\ref{lumflares}, log( L$_{u,\textrm{flare}}$) [erg s$^{-1}$] spans 29.4-31.6 in M0-M1, 28.5-30.5 in M2-M3, and 28-30.5 M4-M6. While the \cwdm sample is sparse, we can already see that when binned by spectral type, the flare luminosities fall within the energies spanned by the field dM sample, with one exception in the late-type bin. Clearly, a larger sample size is needed for a more robust comparison, but preliminary evidence suggests that mid-to-late spectral type dMs in close binary systems may have systematically higher flaring luminosities. Again, more data are necessary to determine the underlying \cwdm flaring luminosity distribution.

Our results show early-to-mid spectral type dMs with close binary companions flare as much as a few hundred times more frequently than field dMs, albeit at slightly higher or similar energies. With enhanced flares, planets will be exposed to higher-levels of X-ray and extreme ultraviolet (EUV) radiation (XUV), which can greatly affect the composition, size, and even presence of a planetary atmosphere. XUV radiation can heat and ionize the upper atmosphere of a planet, leading to atmospheric evaporation from the atmospheric ions being picked up and carried away by stellar wind plasmas \citep{Lammer2007}. If any EUV flux reaches the planetary surface then it can also damage or destroy DNA \citep{Scalo2007}. In addition to the XUV radiation, we expect coronal mass ejections (CMEs) to accompany the largest flares, analogous to what is seen in the Sun \citep[e.g.,][]{Aarnio2011}. Energetic particles from CMEs can compress the magnetosphere, exposing more of the neutral atmosphere which can then be ionized and carried away by incident plasma from the CME. The compression of the magnetosphere by the CME will also allow more UV photons to penetrate deeper into the atmosphere and possibly reach the planetary surface \cite[e.g.,][]{Khodachenko2007}. \cite{Lammer2007} predicts that any Earth-like planet in the HZ of an dM (between 0.05-0.20 au) would lose much of its atmosphere when exposed to XUV flux values 70-100 times to that of the present values at Earth. If the planet did not possess a strong magnetic moment, as suggested for tidally locked planets at distances $<$ 0.20 au \citep{Griesmeier2004, Griesmeier2005, Griesmeier2009}, XUV fluxes $\le$ 50 times that found at Earth would result in the complete loss of the atmosphere within 1 Gyr. The quiescent XUV flux for a dM HZ planet is likely an order of magnitude less than that at Earth, but the flux will increase to 10-100 times that at Earth during dM flares \citep{Scalo2007}.

On the other hand, studies have shown that stellar flares do not pose a threat for habitability for Earth-like planets in the HZ around even the most active flaring dM known \citep[AD Leo;][]{Segura2010}. Much of the damaging EUV flux goes towards photolyzing the ozone layer and never reaches the surface. The corresponding temperature variations in the ozone layer are expected to be small and should have a negligible effect on the surface temperatures of the planet \citep{Segura2010}. However, \cite{Segura2010} posits that if the flares occur more frequently than it takes the atmosphere to equilibrate ($\sim$4 months in their simulation), then more life-damaging UV radiation will be able to reach the planetary surface.

With our reported enhanced flare rates in M dwarfs with close companions, along with reported enhanced activity lifetimes among close binaries \citep{Morgan2012}, the high XUV-radiation environments around dMs is certainly higher, and sustained for longer, than seen around field dMs. This enhancement may have an effect on the habitability of attending exoplanets. Whether or not the XUV fluxes reach levels \citep[70-100 times that seen at Earth;][]{Lammer2007} required to catastrophically disrupt atmospheres of planets in the HZ, requires a more detailed analysis. In addition, our analysis cannot determine whether or not the frequency of flares is high enough to keep a planetary atmosphere out of equilibrium long enough for EUV radiation to reach the planetary surface. Determining the true habitability of exoplanets around dMs is a complex problem that depends on more than just the high-radiation environment imposed by stellar flares. However, more frequent flares in dMs in close binaries certainly may hurt the formation of life, at least, as we know it on Earth.

\section{Summary}\label{Sec:Summary}
Using time-domain data from the Stripe 82 Survey in SDSS, we search for flares around close \cwdm binary pairs. We find nine total flares in eight systems, two of which are magnetically active (as measured by H$\alpha$ emission). We compare these results to the field dM study by K09, in which 271 flares were seen in 236 dM stars. 99 of the field stars had spectroscopic follow-up, with 91 of them identified as magnetically active. For the \cwdm pairs, this corresponds to a flaring frequency of 0.09$\pm$0.03\% (all) and 0.05$\pm$0.03\% (active) and for the dMs a flaring frequency of 0.0108$\pm$0.0007\% (all) and 0.28$\pm$0.05\% (active).

We present preliminary evidence showing that early-type \cwdm pairs (M0-M1) flare $\sim$220 times more frequently then field early-type dMs, and $\sim$15 times more frequently when considering only the active early-type \cwdms and field dMs. Mid-type \cwdms (M2-M3) flare 6.5 times more frequently than mid-type field dMs, but the active mid-type \cwdms appear to flare at approximately the same frequency as the active mid-type field dMs, to within the binomial errors. However, at late spectral types (M4-M6), the \cwdm pairs flare 50\% as frequently as the field dMs and the active \cwdm pairs flare only 8\% as frequently as the active, late-type field dMs. This discrepancy may be attributed to the small size of our sample or the difficulty in finding the intrinsically faint late-type dMs with close WD companions. Perhaps most interesting, is that only two of the eights stars in our sample were shown to be magnetically active compared to 91 of 99 stars in the field dM sample. Assuming that 8\% of flaring stars are inactive (K09), there is a 2.3\% probability that we pull 6 inactive stars from 8 draws. Therefore, we suspect the relatively high number of inactive stars in our flaring sample is due to physical mechanisms involving a close binary companion, rather than small number statistics.

M dwarf flares are one of the main contributors to the Galactic and extragalactic transient background. We considered the contribution to this background by flaring \cwdm pairs and found that the contribution is likely negligible. The average Galactic flare rate was estimated by K09 to be 1.3 flares hr$^{-1}$ deg$^{-2}$ while we estimate a lower-limit for the \cwdm flare rate to be 0.04 flares hr$^{-1}$ deg$^{-2}$ (both for flares $\Delta${\it u} $\ge$ 0.7 and {\it u} $<$ 22), only 3\% of the flares contributed by field dMs.  We re-iterate that since the Stripe 82 dataset is not sensitive to flare duration, we are reporting the likelihood of seeing the \cwdm pairs in a flaring state, rather than a true flare rate.

However, we can use the \cwdms in this study as proxies for dMs with any close companions, not just WDs. We estimate a flare rate for dMs with close companions by normalizing the \cwdm flare rate by the number of \cwdms searched in this study (181). The flare rate for dMs with close companions is then (2.4 $\pm$ 0.8)$\times$10$^{-4}$ flares deg$^{-2}$ hr$^{-1}$ system$^{-1}$ compared to the field dM study flare rate of (2.7 $\pm$ 0.2)$\times$10$^{-5}$ flares deg$^{-2}$ hr$^{-1}$ system$^{-1}$, after a similar normalization (50130 dMs). M dwarfs with close companions flare 6.5 times more frequently than fields dMs.

Previous studies estimate that dM multiplicity at close separations (1$-$100 au) may be as high as 20\% \citep[e.g.,][]{Janson2012, Janson2014, Ward-Duong2015}. Many of these companions are expected to be low-to-equal mass and the close binary system would masquerade as single dMs. Realizing that K09 reported single dM flare rate would also include dMs with unresolved close binary companions, we derive a single dM flare rate using our \cwdm binary flare rate as a proxy. We find that dMs with close binary companions likely dominate the measured flare rate. In the context of SDSS, this makes some sense because the field dM sample is biased towards older, farther away systems, stars on which we would not expect to see many flares. There could be a subset of dMs with close companions, causing them to remain active and flare more often than usual. We hope our results motivate future studies to investigate flare rates for both single and close binary dMs.

Our results also inform the true habitability of attending circumbinary planets. We show that early-type dMs with close companions are $\sim$200 times more likely to flare at energies similar, or slightly more energetic than observed in field dMs. While late-type dMs with close companions show similar flaring fractions and flaring energies as the field dM population. Given our results, we expect the XUV radiation environments to be higher than seen in field dMs, whether or not the XUV radiation reaches high-enough levels to cause significant atmospheric evaporation remains to be seen. In addition, despite the increase in flare frequency, we do not expect these frequencies to reach the levels of some of the most active stars observed like AD Leo \citep{Segura2010}, which studies predict should have no impact on the habitability of an Earth-like planet. We conclude that our reported increase in flares makes the habitability of any attending circumbinary planets less promising than in field dM systems.

One of the consequences of this study is the realization that previously reported single dM flare rates likely include dMs with undetected low-mass companions. These low-mass companions are likely altering the activity history and allowing dMs to remain active for longer and thus, flare at rates higher than what is expected from the field population. We have motivated the need to independently constrain flare rates around field dM and dMs with close binary companions. As a part of these future studies, the following parameters will need to be better constrained:
\begin{enumerate}
	\item The close binary fraction of dMs. There are many studies examining the binarity of dMs as a function of separation \citep[e.g.,][]{Fischer1992, Janson2012, Janson2014,Ward-Duong2015}. Unfortunately, many of these studies are limited to separations $>$1 au projected linear physical separation. One potential solution for probing the closest separations ($<$1 au) is to conduct a high resolution spectroscopic radial velocity survey of nearby, bright dMs where orbital motions can be detected in spectral features.
	\item Flare rates for close binaries remains poorly constrained as our sample size is small and not representative of the population of dMs with close binary companions. 
	\item The physical conditions that give rise to activity enhancement, i.e. the prevention of stellar spin down in the dM, across binary companion mass and system separation is not well constrained. Empirical evidence suggests that dM activity enhancement occurs with a $\sim$0.6 M$_{\odot}$ companion as far out as 1$-$100 au \citep{Morgan2012} and a $>1$M$_{Jup}$ planet within separations of $\sim$0.03 au \citep[e.g.,][]{Poppenhaeger2014}. The parameter space of how companion mass and physical separation affects the activity enhancement in dMs remains incomplete.
\end{enumerate}

Our study addresses the importance of multiplicity (specifically at close separations $<$1 au) in understanding magnetic activity and flaring properties in dMs. Current and future time-domain surveys are ideal laboratories to overcome many of the limitations addressed above. The \cwdm flare rate can be better constrained using nearly all-sky surveys like Lincoln Near-Earth Asteroid Research \citep[Linear;][]{Stokes1998} and Catalina Real-Time Survey \citep[CRTS;][]{Drake2009} that offer hundreds of epochs, albeit without spectral filters. As such, these surveys will offer excellent time-coverage of flaring events, yet, will require overlapping surveys with color-information (such as SDSS) to construct samples of dMs and \cwdms with which to compare flare rates self-similarly. Taking advantage of the high-cadence coverage of {\it Kepler}, we can learn more about the differences in the flare frequency distribution between field dMs and WD+dM. Pan-STARRS and LSST will offer nearly full sky-coverage across many filters, allowing for the construction of large statistical samples of field dMs and WD+dM, which can be used to constrain flare rates and flare energies.

\acknowledgments
The authors would first like to thank the anonymous referee for their extremely helpful comments which greatly improved the quality of this study. The authors would like to thank Chris Theissen, Saurav Dhital, Julie Skinner, Aurora Kesseli, Suzanne Hawley, and Adam Kowalski for fruitful conversations and motivation. A.A.W+D.P.M acknowledge funding from NSF grants AST-1109273 and AST-1255568 and the support from the Research Corporation for Science Advancement's Cottrell Scholarship.

Funding for SDSS-III has been provided by the Alfred P. Sloan Foundation, the Participating Institutions, the National Science Foundation, and the U.S. Department of Energy Office of Science. The SDSS-III web site is http://www.sdss3.org/.

SDSS-III is managed by the Astrophysical Research Consortium for the Participating Institutions of the SDSS-III Collaboration including the University of Arizona, the Brazilian Participation Group, Brookhaven National Laboratory, Carnegie Mellon University, University of Florida, the French Participation Group, the German Participation Group, Harvard University, the Instituto de Astrofisica de Canarias, the Michigan State/Notre Dame/JINA Participation Group, Johns Hopkins University, Lawrence Berkeley National Laboratory, Max Planck Institute for Astrophysics, Max Planck Institute for Extraterrestrial Physics, New Mexico State University, New York University, Ohio State University, Pennsylvania State University, University of Portsmouth, Princeton University, the Spanish Participation Group, University of Tokyo, University of Utah, Vanderbilt University, University of Virginia, University of Washington, and Yale University.

\bibliography{ms}

\begin{thebibliography}{}
\expandafter\ifx\csname natexlab\endcsname\relax\def\natexlab#1{#1}\fi

\bibitem[{{Aarnio} {et~al.}(2011){Aarnio}, {Stassun}, {Hughes}, \&
  {McGregor}}]{Aarnio2011}
{Aarnio}, A.~N., {Stassun}, K.~G., {Hughes}, W.~J., \& {McGregor}, S.~L. 2011,
  \solphys, 268, 195

\bibitem[{{Abazajian} {et~al.}(2009){Abazajian}, {Adelman-McCarthy},
  {Ag{\"u}eros}, {Allam}, {Allende Prieto}, {An}, {Anderson}, {Anderson},
  {Annis}, {Bahcall}, \& et~al.}]{Abazajian2009}
{Abazajian}, K.~N., {Adelman-McCarthy}, J.~K., {Ag{\"u}eros}, M.~A., {et~al.}
  2009, \apjs, 182, 543

\bibitem[{{Alam} {et~al.}(2015){Alam}, {Albareti}, {Allende Prieto}, {Anders},
  {Anderson}, {Andrews}, {Armengaud}, {Aubourg}, {Bailey}, {Bautista}, \&
  et~al.}]{Alam2015}
{Alam}, S., {Albareti}, F.~D., {Allende Prieto}, C., {et~al.} 2015, ArXiv
  e-prints, arXiv:1501.00963

\bibitem[{{Audard} {et~al.}(2000){Audard}, {G{\"u}del}, {Drake}, \&
  {Kashyap}}]{Audard2000}
{Audard}, M., {G{\"u}del}, M., {Drake}, J.~J., \& {Kashyap}, V.~L. 2000, \apj,
  541, 396

\bibitem[{{Ballard} \& {Johnson}(2014)}]{Ballard2014}
{Ballard}, S., \& {Johnson}, J.~A. 2014, ArXiv e-prints, arXiv:1410.4192

\bibitem[{{Becker} {et~al.}(2004){Becker}, {Wittman}, {Boeshaar},
  {Clocchiatti}, {Dell'Antonio}, {Frail}, {Halpern}, {Margoniner}, {Norman},
  {Tyson}, \& {Schommer}}]{Becker2004}
{Becker}, A.~C., {Wittman}, D.~M., {Boeshaar}, P.~C., {et~al.} 2004, \apj, 611,
  418

\bibitem[{{Berger} {et~al.}(2013){Berger}, {Leibler}, {Chornock}, {Rest},
  {Foley}, {Soderberg}, {Price}, {Burgett}, {Chambers}, {Flewelling}, {Huber},
  {Magnier}, {Metcalfe}, {Stubbs}, \& {Tonry}}]{Berger2013}
{Berger}, E., {Leibler}, C.~N., {Chornock}, R., {et~al.} 2013, \apj, 779, 18

\bibitem[{{Berta-Thompson} {et~al.}(2015){Berta-Thompson}, {Irwin},
  {Charbonneau}, {Newton}, {Dittmann}, {Astudillo-Defru}, {Bonfils}, {Gillon},
  {Jehin}, {Stark}, {Stalder}, {Bouchy}, {Delfosse}, {Forveille}, {Lovis},
  {Mayor}, {Neves}, {Pepe}, {Santos}, {Udry}, \&
  {W{\"u}nsche}}]{Berta-Thompson2015}
{Berta-Thompson}, Z.~K., {Irwin}, J., {Charbonneau}, D., {et~al.} 2015, ArXiv
  e-prints, arXiv:1511.03550

\bibitem[{{Bochanski} {et~al.}(2010){Bochanski}, {Hawley}, {Covey}, {West},
  {Reid}, {Golimowski}, \& {Ivezi{\'c}}}]{Bochanski2010}
{Bochanski}, J.~J., {Hawley}, S.~L., {Covey}, K.~R., {et~al.} 2010, \aj, 139,
  2679

\bibitem[{{Bochanski} {et~al.}(2011){Bochanski}, {Hawley}, \&
  {West}}]{Bochanski2011}
{Bochanski}, J.~J., {Hawley}, S.~L., \& {West}, A.~A. 2011, \aj, 141, 98

\bibitem[{{Bochanski}(2008)}]{Bochanski2008}
{Bochanski}, Jr., J.~J. 2008, PhD thesis, University of Washington

\bibitem[{{Catal{\'a}n} {et~al.}(2008){Catal{\'a}n}, {Isern},
  {Garc{\'{\i}}a-Berro}, {Ribas}, {Allende Prieto}, \& {Bonanos}}]{Catalan2008}
{Catal{\'a}n}, S., {Isern}, J., {Garc{\'{\i}}a-Berro}, E., {et~al.} 2008, \aap,
  477, 213

\bibitem[{{Chabrier}(2003)}]{Chabrier2003}
{Chabrier}, G. 2003, \pasp, 115, 763

\bibitem[{{Doi} {et~al.}(2010){Doi}, {Tanaka}, {Fukugita}, {Gunn}, {Yasuda},
  {Ivezi{\'c}}, {Brinkmann}, {de Haars}, {Kleinman}, {Krzesinski}, \& {French
  Leger}}]{Doi2010}
{Doi}, M., {Tanaka}, M., {Fukugita}, M., {et~al.} 2010, \aj, 139, 1628

\bibitem[{{Doyle} {et~al.}(2011){Doyle}, {Carter}, {Fabrycky}, {Slawson},
  {Howell}, {Winn}, {Orosz}, {Welsh}, {Quinn}, {Latham}, {Torres}, {Buchhave},
  {Marcy}, {Fortney}, {Shporer}, {Ford}, {Lissauer}, {Ragozzine}, {Rucker},
  {Batalha}, {Jenkins}, {Borucki}, {Koch}, {Middour}, {Hall}, {McCauliff},
  {Fanelli}, {Quintana}, {Holman}, {Caldwell}, {Still}, {Stefanik}, {Brown},
  {Esquerdo}, {Tang}, {Furesz}, {Geary}, {Berlind}, {Calkins}, {Short},
  {Steffen}, {Sasselov}, {Dunham}, {Cochran}, {Boss}, {Haas}, {Buzasi}, \&
  {Fischer}}]{Doyle2011}
{Doyle}, L.~R., {Carter}, J.~A., {Fabrycky}, D.~C., {et~al.} 2011, Science,
  333, 1602

\bibitem[{{Drake} {et~al.}(2009){Drake}, {Djorgovski}, {Mahabal}, {Beshore},
  {Larson}, {Graham}, {Williams}, {Christensen}, {Catelan}, {Boattini},
  {Gibbs}, {Hill}, \& {Kowalski}}]{Drake2009}
{Drake}, A.~J., {Djorgovski}, S.~G., {Mahabal}, A., {et~al.} 2009, \apj, 696,
  870

\bibitem[{{Dressing} \& {Charbonneau}(2013)}]{Dressing2013}
{Dressing}, C.~D., \& {Charbonneau}, D. 2013, \apj, 767, 95

\bibitem[{{Dressing} \& {Charbonneau}(2015)}]{Dressing2015a}
---. 2015, ArXiv e-prints, arXiv:1501.01623

\bibitem[{{Eason} {et~al.}(1992){Eason}, {Giampapa}, {Radick}, {Worden}, \&
  {Hege}}]{Eason1992}
{Eason}, E.~L.~E., {Giampapa}, M.~S., {Radick}, R.~R., {Worden}, S.~P., \&
  {Hege}, E.~K. 1992, \aj, 104, 1161

\bibitem[{{Fischer} \& {Marcy}(1992)}]{Fischer1992}
{Fischer}, D.~A., \& {Marcy}, G.~W. 1992, \apj, 396, 178

\bibitem[{{Frieman} {et~al.}(2008){Frieman}, {Bassett}, {Becker}, {Choi},
  {Cinabro}, {DeJongh}, {Depoy}, {Dilday}, {Doi}, {Garnavich}, {Hogan},
  {Holtzman}, {Im}, {Jha}, {Kessler}, {Konishi}, {Lampeitl}, {Marriner},
  {Marshall}, {McGinnis}, {Miknaitis}, {Nichol}, {Prieto}, {Riess}, {Richmond},
  {Romani}, {Sako}, {Schneider}, {Smith}, {Takanashi}, {Tokita}, {van der
  Heyden}, {Yasuda}, {Zheng}, {Adelman-McCarthy}, {Annis}, {Assef},
  {Barentine}, {Bender}, {Blandford}, {Boroski}, {Bremer}, {Brewington},
  {Collins}, {Crotts}, {Dembicky}, {Eastman}, {Edge}, {Edmondson}, {Elson},
  {Eyler}, {Filippenko}, {Foley}, {Frank}, {Goobar}, {Gueth}, {Gunn},
  {Harvanek}, {Hopp}, {Ihara}, {Ivezi{\'c}}, {Kahn}, {Kaplan}, {Kent},
  {Ketzeback}, {Kleinman}, {Kollatschny}, {Kron}, {Krzesi{\'n}ski}, {Lamenti},
  {Leloudas}, {Lin}, {Long}, {Lucey}, {Lupton}, {Malanushenko}, {Malanushenko},
  {McMillan}, {Mendez}, {Morgan}, {Morokuma}, {Nitta}, {Ostman}, {Pan},
  {Rockosi}, {Romer}, {Ruiz-Lapuente}, {Saurage}, {Schlesinger}, {Snedden},
  {Sollerman}, {Stoughton}, {Stritzinger}, {Subba Rao}, {Tucker}, {Vaisanen},
  {Watson}, {Watters}, {Wheeler}, {Yanny}, \& {York}}]{Frieman2008}
{Frieman}, J.~A., {Bassett}, B., {Becker}, A., {et~al.} 2008, \aj, 135, 338

\bibitem[{{Grie{\ss}meier} {et~al.}(2009){Grie{\ss}meier}, {Stadelmann},
  {Grenfell}, {Lammer}, \& {Motschmann}}]{Griesmeier2009}
{Grie{\ss}meier}, J.-M., {Stadelmann}, A., {Grenfell}, J.~L., {Lammer}, H., \&
  {Motschmann}, U. 2009, Icarus, 199, 526

\bibitem[{{Grie{\ss}meier} {et~al.}(2005){Grie{\ss}meier}, {Stadelmann},
  {Lammer}, {Belisheva}, \& {Motschmann}}]{Griesmeier2005}
{Grie{\ss}meier}, J.-M., {Stadelmann}, A., {Lammer}, H., {Belisheva}, N., \&
  {Motschmann}, U. 2005, in ESA Special Publication, Vol. 588, 39TH ESLAB
  Symposium on Trends in Space Science and Cosmic Vision 2020, ed. F.~{Favata},
  J.~{Sanz-Forcada}, A.~{Gim{\'e}nez}, \& B.~{Battrick}, 305

\bibitem[{{Grie{\ss}meier} {et~al.}(2004){Grie{\ss}meier}, {Stadelmann},
  {Penz}, {Lammer}, {Selsis}, {Ribas}, {Guinan}, {Motschmann}, {Biernat}, \&
  {Weiss}}]{Griesmeier2004}
{Grie{\ss}meier}, J.-M., {Stadelmann}, A., {Penz}, T., {et~al.} 2004, \aap,
  425, 753

\bibitem[{{Gu} \& {Ai}(2011)}]{Gu2011}
{Gu}, M.-F., \& {Ai}, Y.~L. 2011, \aap, 528, A95

\bibitem[{{G{\"u}del} {et~al.}(2003){G{\"u}del}, {Audard}, {Kashyap}, {Drake},
  \& {Guinan}}]{Gudel2003}
{G{\"u}del}, M., {Audard}, M., {Kashyap}, V.~L., {Drake}, J.~J., \& {Guinan},
  E.~F. 2003, \apj, 582, 423

\bibitem[{{Gunn} {et~al.}(1998){Gunn}, {Carr}, {Rockosi}, {Sekiguchi}, {Berry},
  {Elms}, {de Haas}, {Ivezi{\'c}}, {Knapp}, {Lupton}, {Pauls}, {Simcoe},
  {Hirsch}, {Sanford}, {Wang}, {York}, {Harris}, {Annis}, {Bartozek},
  {Boroski}, {Bakken}, {Haldeman}, {Kent}, {Holm}, {Holmgren}, {Petravick},
  {Prosapio}, {Rechenmacher}, {Doi}, {Fukugita}, {Shimasaku}, {Okada}, {Hull},
  {Siegmund}, {Mannery}, {Blouke}, {Heidtman}, {Schneider}, {Lucinio}, \&
  {Brinkman}}]{Gunn1998}
{Gunn}, J.~E., {Carr}, M., {Rockosi}, C., {et~al.} 1998, \aj, 116, 3040

\bibitem[{{Gunn} {et~al.}(2006){Gunn}, {Siegmund}, {Mannery}, {Owen}, {Hull},
  {Leger}, {Carey}, {Knapp}, {York}, {Boroski}, {Kent}, {Lupton}, {Rockosi},
  {Evans}, {Waddell}, {Anderson}, {Annis}, {Barentine}, {Bartoszek}, {Bastian},
  {Bracker}, {Brewington}, {Briegel}, {Brinkmann}, {Brown}, {Carr},
  {Czarapata}, {Drennan}, {Dombeck}, {Federwitz}, {Gillespie}, {Gonzales},
  {Hansen}, {Harvanek}, {Hayes}, {Jordan}, {Kinney}, {Klaene}, {Kleinman},
  {Kron}, {Kresinski}, {Lee}, {Limmongkol}, {Lindenmeyer}, {Long}, {Loomis},
  {McGehee}, {Mantsch}, {Neilsen}, {Neswold}, {Newman}, {Nitta}, {Peoples},
  {Pier}, {Prieto}, {Prosapio}, {Rivetta}, {Schneider}, {Snedden}, \&
  {Wang}}]{Gunn2006}
{Gunn}, J.~E., {Siegmund}, W.~A., {Mannery}, E.~J., {et~al.} 2006, \aj, 131,
  2332

\bibitem[{{Hawley} {et~al.}(2014){Hawley}, {Davenport}, {Kowalski},
  {Wisniewski}, {Hebb}, {Deitrick}, \& {Hilton}}]{Hawley2014}
{Hawley}, S.~L., {Davenport}, J.~R.~A., {Kowalski}, A.~F., {et~al.} 2014, \apj,
  797, 121

\bibitem[{{Hawley} \& {Pettersen}(1991)}]{Hawley1991}
{Hawley}, S.~L., \& {Pettersen}, B.~R. 1991, \apj, 378, 725

\bibitem[{{Hawley} {et~al.}(2007){Hawley}, {Walkowicz}, {Allred}, \&
  {Valenti}}]{Hawley2007}
{Hawley}, S.~L., {Walkowicz}, L.~M., {Allred}, J.~C., \& {Valenti}, J.~A. 2007,
  \pasp, 119, 67

\bibitem[{{Henry} {et~al.}(1994){Henry}, {Kirkpatrick}, \&
  {Simons}}]{Henry1994}
{Henry}, T.~J., {Kirkpatrick}, J.~D., \& {Simons}, D.~A. 1994, \aj, 108, 1437

\bibitem[{{Hilton}(2011)}]{Hilton2011}
{Hilton}, E.~J. 2011, PhD thesis, University of Washington

\bibitem[{{Hilton} {et~al.}(2010){Hilton}, {West}, {Hawley}, \&
  {Kowalski}}]{Hilton2010}
{Hilton}, E.~J., {West}, A.~A., {Hawley}, S.~L., \& {Kowalski}, A.~F. 2010,
  \aj, 140, 1402

\bibitem[{{Hjellming} \& {Taam}(1991)}]{Hjellming1991}
{Hjellming}, M.~S., \& {Taam}, R.~E. 1991, \apj, 370, 709

\bibitem[{{Ivanova} {et~al.}(2013){Ivanova}, {Justham}, {Chen}, {De Marco},
  {Fryer}, {Gaburov}, {Ge}, {Glebbeek}, {Han}, {Li}, {Lu}, {Marsh},
  {Podsiadlowski}, {Potter}, {Soker}, {Taam}, {Tauris}, {van den Heuvel}, \&
  {Webbink}}]{Ivanova2013}
{Ivanova}, N., {Justham}, S., {Chen}, X., {et~al.} 2013, \aapr, 21, 59

\bibitem[{{Ivezi{\'c}} {et~al.}(2004){Ivezi{\'c}}, {Lupton}, {Schlegel},
  {Boroski}, {Adelman-McCarthy}, {Yanny}, {Kent}, {Stoughton}, {Finkbeiner},
  {Padmanabhan}, {Rockosi}, {Gunn}, {Knapp}, {Strauss}, {Richards},
  {Eisenstein}, {Nicinski}, {Kleinman}, {Krzesinski}, {Newman}, {Snedden},
  {Thakar}, {Szalay}, {Munn}, {Smith}, {Tucker}, \& {Lee}}]{Ivezic2004}
{Ivezi{\'c}}, {\v Z}., {Lupton}, R.~H., {Schlegel}, D., {et~al.} 2004,
  Astronomische Nachrichten, 325, 583

\bibitem[{{Ivezi{\'c}} {et~al.}(2007){Ivezi{\'c}}, {Smith}, {Miknaitis}, {Lin},
  {Tucker}, {Lupton}, {Gunn}, {Knapp}, {Strauss}, {Sesar}, {Doi}, {Tanaka},
  {Fukugita}, {Holtzman}, {Kent}, {Yanny}, {Schlegel}, {Finkbeiner},
  {Padmanabhan}, {Rockosi}, {Juri{\'c}}, {Bond}, {Lee}, {Stoughton}, {Jester},
  {Harris}, {Harding}, {Morrison}, {Brinkmann}, {Schneider}, \&
  {York}}]{Ivezic2007}
{Ivezi{\'c}}, {\v Z}., {Smith}, J.~A., {Miknaitis}, G., {et~al.} 2007, \aj,
  134, 973

\bibitem[{{Janson} {et~al.}(2014){Janson}, {Bergfors}, {Brandner},
  {Kudryavtseva}, {Hormuth}, {Hippler}, \& {Henning}}]{Janson2014}
{Janson}, M., {Bergfors}, C., {Brandner}, W., {et~al.} 2014, \apj, 789, 102

\bibitem[{{Janson} {et~al.}(2012){Janson}, {Hormuth}, {Bergfors}, {Brandner},
  {Hippler}, {Daemgen}, {Kudryavtseva}, {Schmalzl}, {Schnupp}, \&
  {Henning}}]{Janson2012}
{Janson}, M., {Hormuth}, F., {Bergfors}, C., {et~al.} 2012, \apj, 754, 44

\bibitem[{{Kaiser}(2004)}]{Kaiser2004}
{Kaiser}, N. 2004, in Society of Photo-Optical Instrumentation Engineers (SPIE)
  Conference Series, Vol. 5489, Ground-based Telescopes, ed. J.~M. {Oschmann},
  Jr., 11--22

\bibitem[{{Kepler} {et~al.}(2007){Kepler}, {Kleinman}, {Nitta}, {Koester},
  {Castanheira}, {Giovannini}, {Costa}, \& {Althaus}}]{Kepler2007}
{Kepler}, S.~O., {Kleinman}, S.~J., {Nitta}, A., {et~al.} 2007, \mnras, 375,
  1315

\bibitem[{{Khodachenko} {et~al.}(2007){Khodachenko}, {Ribas}, {Lammer},
  {Grie{\ss}meier}, {Leitner}, {Selsis}, {Eiroa}, {Hanslmeier}, {Biernat},
  {Farrugia}, \& {Rucker}}]{Khodachenko2007}
{Khodachenko}, M.~L., {Ribas}, I., {Lammer}, H., {et~al.} 2007, Astrobiology,
  7, 167

\bibitem[{{Koenigl}(1991)}]{Koenigl1991}
{Koenigl}, A. 1991, \apjl, 370, L39

\bibitem[{{Koester} {et~al.}(2001){Koester}, {Napiwotzki}, {Christlieb},
  {Drechsel}, {Hagen}, {Heber}, {Homeier}, {Karl}, {Leibundgut}, {Moehler},
  {Nelemans}, {Pauli}, {Reimers}, {Renzini}, \& {Yungelson}}]{Koester2001}
{Koester}, D., {Napiwotzki}, R., {Christlieb}, N., {et~al.} 2001, \aap, 378,
  556

\bibitem[{{Korycansky} \& {Papaloizou}(1995)}]{Korycansky1995}
{Korycansky}, D.~G., \& {Papaloizou}, J.~C.~B. 1995, \mnras, 274, 85

\bibitem[{{Kostov} {et~al.}(2013){Kostov}, {McCullough}, {Hinse}, {Tsvetanov},
  {H{\'e}brard}, {D{\'{\i}}az}, {Deleuil}, \& {Valenti}}]{Kostov2013}
{Kostov}, V.~B., {McCullough}, P.~R., {Hinse}, T.~C., {et~al.} 2013, \apj, 770,
  52

\bibitem[{{Kostrzewa-Rutkowska} {et~al.}(2013){Kostrzewa-Rutkowska},
  {Koz{\l}owski}, {Wyrzykowski}, {Djorgovski}, {Glikman}, {Mahabal}, \&
  {Koposov}}]{Kostrzewa-Rutkowska2013}
{Kostrzewa-Rutkowska}, Z., {Koz{\l}owski}, S., {Wyrzykowski}, {\L}., {et~al.}
  2013, \apj, 778, 168

\bibitem[{{Kowalski} {et~al.}(2014){Kowalski}, {Cauzzi}, \&
  {Fletcher}}]{Kowalski2014}
{Kowalski}, A.~F., {Cauzzi}, G., \& {Fletcher}, L. 2014, ArXiv e-prints,
  arXiv:1411.0770

\bibitem[{{Kowalski} {et~al.}(2009){Kowalski}, {Hawley}, {Hilton}, {Becker},
  {West}, {Bochanski}, \& {Sesar}}]{Kowalski2009}
{Kowalski}, A.~F., {Hawley}, S.~L., {Hilton}, E.~J., {et~al.} 2009, \aj, 138,
  633

\bibitem[{{Kowalski} {et~al.}(2010){Kowalski}, {Hawley}, {Holtzman},
  {Wisniewski}, \& {Hilton}}]{Kowalski2010}
{Kowalski}, A.~F., {Hawley}, S.~L., {Holtzman}, J.~A., {Wisniewski}, J.~P., \&
  {Hilton}, E.~J. 2010, \apjl, 714, L98

\bibitem[{{Kulkarni} \& {Rau}(2006)}]{Kulkarni2006}
{Kulkarni}, S.~R., \& {Rau}, A. 2006, \apjl, 644, L63

\bibitem[{{Lacy} {et~al.}(1976){Lacy}, {Moffett}, \& {Evans}}]{Lacy1976}
{Lacy}, C.~H., {Moffett}, T.~J., \& {Evans}, D.~S. 1976, \apjs, 30, 85

\bibitem[{{Lammer} {et~al.}(2007){Lammer}, {Lichtenegger}, {Kulikov},
  {Grie{\ss}meier}, {Terada}, {Erkaev}, {Biernat}, {Khodachenko}, {Ribas},
  {Penz}, \& {Selsis}}]{Lammer2007}
{Lammer}, H., {Lichtenegger}, H.~I.~M., {Kulikov}, Y.~N., {et~al.} 2007,
  Astrobiology, 7, 185

\bibitem[{{LSST Science Collaboration} {et~al.}(2009){LSST Science
  Collaboration}, {Abell}, {Allison}, {Anderson}, {Andrew}, {Angel}, {Armus},
  {Arnett}, {Asztalos}, {Axelrod}, \& et~al.}]{LSST2009}
{LSST Science Collaboration}, {Abell}, P.~A., {Allison}, J., {et~al.} 2009,
  ArXiv e-prints, arXiv:0912.0201

\bibitem[{{Martin} {et~al.}(2005){Martin}, {Fanson}, {Schiminovich},
  {Morrissey}, {Friedman}, {Barlow}, {Conrow}, {Grange}, {Jelinsky},
  {Milliard}, {Siegmund}, {Bianchi}, {Byun}, {Donas}, {Forster}, {Heckman},
  {Lee}, {Madore}, {Malina}, {Neff}, {Rich}, {Small}, {Surber}, {Szalay},
  {Welsh}, \& {Wyder}}]{Martin2005}
{Martin}, D.~C., {Fanson}, J., {Schiminovich}, D., {et~al.} 2005, \apjl, 619,
  L1

\bibitem[{{Matt} {et~al.}(2012){Matt}, {Pinz{\'o}n}, {Greene}, \&
  {Pudritz}}]{Matt2012}
{Matt}, S.~P., {Pinz{\'o}n}, G., {Greene}, T.~P., \& {Pudritz}, R.~E. 2012,
  \apj, 745, 101

\bibitem[{{Meibom} \& {Mathieu}(2005)}]{Meibom2005}
{Meibom}, S., \& {Mathieu}, R.~D. 2005, \apj, 620, 970

\bibitem[{{Moffett}(1974)}]{Moffett1974}
{Moffett}, T.~J. 1974, \apjs, 29, 1

\bibitem[{{Morgan} {et~al.}(2012){Morgan}, {West}, {Garc{\'e}s}, {Catal{\'a}n},
  {Dhital}, {Fuchs}, \& {Silvestri}}]{Morgan2012}
{Morgan}, D.~P., {West}, A.~A., {Garc{\'e}s}, A., {et~al.} 2012, \aj, 144, 93

\bibitem[{{Orosz} {et~al.}(2012{\natexlab{a}}){Orosz}, {Welsh}, {Carter},
  {Fabrycky}, {Cochran}, {Endl}, {Ford}, {Haghighipour}, {MacQueen}, {Mazeh},
  {Sanchis-Ojeda}, {Short}, {Torres}, {Agol}, {Buchhave}, {Doyle}, {Isaacson},
  {Lissauer}, {Marcy}, {Shporer}, {Windmiller}, {Barclay}, {Boss}, {Clarke},
  {Fortney}, {Geary}, {Holman}, {Huber}, {Jenkins}, {Kinemuchi}, {Kruse},
  {Ragozzine}, {Sasselov}, {Still}, {Tenenbaum}, {Uddin}, {Winn}, {Koch}, \&
  {Borucki}}]{Orosz2012}
{Orosz}, J.~A., {Welsh}, W.~F., {Carter}, J.~A., {et~al.} 2012{\natexlab{a}},
  Science, 337, 1511

\bibitem[{{Orosz} {et~al.}(2012{\natexlab{b}}){Orosz}, {Welsh}, {Carter},
  {Brugamyer}, {Buchhave}, {Cochran}, {Endl}, {Ford}, {MacQueen}, {Short},
  {Torres}, {Windmiller}, {Agol}, {Barclay}, {Caldwell}, {Clarke}, {Doyle},
  {Fabrycky}, {Geary}, {Haghighipour}, {Holman}, {Ibrahim}, {Jenkins},
  {Kinemuchi}, {Li}, {Lissauer}, {Pr{\v s}a}, {Ragozzine}, {Shporer}, {Still},
  \& {Wade}}]{Orosz2012a}
---. 2012{\natexlab{b}}, \apj, 758, 87

\bibitem[{{Osten} \& {Bastian}(2008)}]{Osten2008}
{Osten}, R.~A., \& {Bastian}, T.~S. 2008, \apj, 674, 1078

\bibitem[{{Osten} {et~al.}(2010){Osten}, {Godet}, {Drake}, {Tueller},
  {Cummings}, {Krimm}, {Pye}, {Pal'shin}, {Golenetskii}, {Reale}, {Oates},
  {Page}, \& {Melandri}}]{Osten2010}
{Osten}, R.~A., {Godet}, O., {Drake}, S., {et~al.} 2010, \apj, 721, 785

\bibitem[{{Palanque-Delabrouille} {et~al.}(2011){Palanque-Delabrouille},
  {Yeche}, {Myers}, {Petitjean}, {Ross}, {Sheldon}, {Aubourg}, {Delubac}, {Le
  Goff}, {P{\^a}ris}, {Rich}, {Dawson}, {Schneider}, \&
  {Weaver}}]{Palanque-Delabrouille2011}
{Palanque-Delabrouille}, N., {Yeche}, C., {Myers}, A.~D., {et~al.} 2011, \aap,
  530, A122

\bibitem[{{Papaloizou} \& {Terquem}(1997)}]{Papaloizou1997}
{Papaloizou}, J.~C.~B., \& {Terquem}, C. 1997, \mnras, 287, 771

\bibitem[{{Pier} {et~al.}(2003){Pier}, {Munn}, {Hindsley}, {Hennessy}, {Kent},
  {Lupton}, \& {Ivezi{\'c}}}]{Pier2003}
{Pier}, J.~R., {Munn}, J.~A., {Hindsley}, R.~B., {et~al.} 2003, \aj, 125, 1559

\bibitem[{{Poppenhaeger}(2011)}]{Poppenhaeger2011a}
{Poppenhaeger}, K. 2011, in Astronomical Society of the Pacific Conference
  Series, Vol. 448, 16th Cambridge Workshop on Cool Stars, Stellar Systems, and
  the Sun, ed. C.~{Johns-Krull}, M.~K. {Browning}, \& A.~A. {West}, 1225

\bibitem[{{Poppenhaeger} \& {Schmitt}(2011)}]{Poppenhaeger2011}
{Poppenhaeger}, K., \& {Schmitt}, J.~H.~M.~M. 2011, \apj, 735, 59

\bibitem[{{Poppenhaeger} \& {Wolk}(2014)}]{Poppenhaeger2014}
{Poppenhaeger}, K., \& {Wolk}, S.~J. 2014, \aap, 565, L1

\bibitem[{{Rau} {et~al.}(2008){Rau}, {Ofek}, {Kulkarni}, {Madore}, {Pevunova},
  \& {Ajello}}]{Rau2008}
{Rau}, A., {Ofek}, E.~O., {Kulkarni}, S.~R., {et~al.} 2008, \apj, 682, 1205

\bibitem[{{Rebassa-Mansergas} {et~al.}(2013){Rebassa-Mansergas},
  {Agurto-Gangas}, {Schreiber}, {G{\"a}nsicke}, \&
  {Koester}}]{Rebassa-Mansergas2013}
{Rebassa-Mansergas}, A., {Agurto-Gangas}, C., {Schreiber}, M.~R.,
  {G{\"a}nsicke}, B.~T., \& {Koester}, D. 2013, \mnras, 433, 3398

\bibitem[{{Rebassa-Mansergas} {et~al.}(2012){Rebassa-Mansergas}, {Nebot
  G{\'o}mez-Mor{\'a}n}, {Schreiber}, {G{\"a}nsicke}, {Schwope}, {Gallardo}, \&
  {Koester}}]{Rebassa-Mansergas2012}
{Rebassa-Mansergas}, A., {Nebot G{\'o}mez-Mor{\'a}n}, A., {Schreiber}, M.~R.,
  {et~al.} 2012, \mnras, 419, 806

\bibitem[{{Rebassa-Mansergas} {et~al.}(2014){Rebassa-Mansergas}, {Parsons},
  {Copperwheat}, {Justham}, {G{\"a}nsicke}, {Schreiber}, {Marsh}, \&
  {Dhillon}}]{Rebassa-Mansergas2014}
{Rebassa-Mansergas}, A., {Parsons}, S.~G., {Copperwheat}, C.~M., {et~al.} 2014,
  \apj, 790, 28

\bibitem[{{Reid} \& {Hawley}(2005)}]{Reid2005}
{Reid}, I.~N., \& {Hawley}, S.~L. 2005, {New light on dark stars : red dwarfs,
  low-mass stars, brown dwarfs}, doi:10.1007/3-540-27610-6

\bibitem[{{Reid} {et~al.}(2004){Reid}, {Cruz}, {Allen}, {Mungall}, {Kilkenny},
  {Liebert}, {Hawley}, {Fraser}, {Covey}, {Lowrance}, {Kirkpatrick}, \&
  {Burgasser}}]{Reid2004}
{Reid}, I.~N., {Cruz}, K.~L., {Allen}, P., {et~al.} 2004, \aj, 128, 463

\bibitem[{{Ricker} {et~al.}(2014){Ricker}, {Winn}, {Vanderspek}, {Latham},
  {Bakos}, {Bean}, {Berta-Thompson}, {Brown}, {Buchhave}, {Butler}, {Butler},
  {Chaplin}, {Charbonneau}, {Christensen-Dalsgaard}, {Clampin}, {Deming},
  {Doty}, {De Lee}, {Dressing}, {Dunham}, {Endl}, {Fressin}, {Ge}, {Henning},
  {Holman}, {Howard}, {Ida}, {Jenkins}, {Jernigan}, {Johnson}, {Kaltenegger},
  {Kawai}, {Kjeldsen}, {Laughlin}, {Levine}, {Lin}, {Lissauer}, {MacQueen},
  {Marcy}, {McCullough}, {Morton}, {Narita}, {Paegert}, {Palle}, {Pepe},
  {Pepper}, {Quirrenbach}, {Rinehart}, {Sasselov}, {Sato}, {Seager},
  {Sozzetti}, {Stassun}, {Sullivan}, {Szentgyorgyi}, {Torres}, {Udry}, \&
  {Villasenor}}]{Ricker2014}
{Ricker}, G.~R., {Winn}, J.~N., {Vanderspek}, R., {et~al.} 2014, in Society of
  Photo-Optical Instrumentation Engineers (SPIE) Conference Series, Vol. 9143,
  Society of Photo-Optical Instrumentation Engineers (SPIE) Conference Series,
  20

\bibitem[{{Robinson} {et~al.}(2005){Robinson}, {Wheatley}, {Welsh}, {Forster},
  {Morrissey}, {Seibert}, {Rich}, {Salim}, {Barlow}, {Bianchi}, {Byun},
  {Donas}, {Friedman}, {Heckman}, {Jelinsky}, {Lee}, {Madore}, {Malina},
  {Martin}, {Milliard}, {Neff}, {Schiminovich}, {Siegmund}, {Small}, {Szalay},
  \& {Wyder}}]{Robinson2005}
{Robinson}, R.~D., {Wheatley}, J.~M., {Welsh}, B.~Y., {et~al.} 2005, \apj, 633,
  447

\bibitem[{{Rykoff} {et~al.}(2005){Rykoff}, {Aharonian}, {Akerlof}, {Alatalo},
  {Ashley}, {G{\"u}ver}, {Horns}, {Kehoe}, {Kizilo{\v g}lu}, {McKay},
  {{\"O}zel}, {Phillips}, {Quimby}, {Schaefer}, {Smith}, {Swan}, {Vestrand},
  {Wheeler}, {Wren}, \& {Yost}}]{Rykoff2005}
{Rykoff}, E.~S., {Aharonian}, F., {Akerlof}, C.~W., {et~al.} 2005, \apj, 631,
  1032

\bibitem[{{Sako} {et~al.}(2008){Sako}, {Bassett}, {Becker}, {Cinabro},
  {DeJongh}, {Depoy}, {Dilday}, {Doi}, {Frieman}, {Garnavich}, {Hogan},
  {Holtzman}, {Jha}, {Kessler}, {Konishi}, {Lampeitl}, {Marriner}, {Miknaitis},
  {Nichol}, {Prieto}, {Riess}, {Richmond}, {Romani}, {Schneider}, {Smith},
  {SubbaRao}, {Takanashi}, {Tokita}, {van der Heyden}, {Yasuda}, {Zheng},
  {Barentine}, {Brewington}, {Choi}, {Dembicky}, {Harnavek}, {Ihara}, {Im},
  {Ketzeback}, {Kleinman}, {Krzesi{\'n}ski}, {Long}, {Malanushenko},
  {Malanushenko}, {McMillan}, {Morokuma}, {Nitta}, {Pan}, {Saurage}, \&
  {Snedden}}]{Sako2008}
{Sako}, M., {Bassett}, B., {Becker}, A., {et~al.} 2008, \aj, 135, 348

\bibitem[{{Scalo} {et~al.}(2007){Scalo}, {Kaltenegger}, {Segura}, {Fridlund},
  {Ribas}, {Kulikov}, {Grenfell}, {Rauer}, {Odert}, {Leitzinger}, {Selsis},
  {Khodachenko}, {Eiroa}, {Kasting}, \& {Lammer}}]{Scalo2007}
{Scalo}, J., {Kaltenegger}, L., {Segura}, A.~G., {et~al.} 2007, Astrobiology,
  7, 85

\bibitem[{{Schmidt} {et~al.}(2012){Schmidt}, {Kowalski}, {Hawley}, {Hilton},
  {Wisniewski}, \& {Tofflemire}}]{Schmidt2012}
{Schmidt}, S.~J., {Kowalski}, A.~F., {Hawley}, S.~L., {et~al.} 2012, \apj, 745,
  14

\bibitem[{{Segura} {et~al.}(2010){Segura}, {Walkowicz}, {Meadows}, {Kasting},
  \& {Hawley}}]{Segura2010}
{Segura}, A., {Walkowicz}, L.~M., {Meadows}, V., {Kasting}, J., \& {Hawley}, S.
  2010, Astrobiology, 10, 751

\bibitem[{{Sesar} {et~al.}(2010){Sesar}, {Ivezi{\'c}}, {Grammer}, {Morgan},
  {Becker}, {Juri{\'c}}, {De Lee}, {Annis}, {Beers}, {Fan}, {Lupton}, {Gunn},
  {Knapp}, {Jiang}, {Jester}, {Johnston}, \& {Lampeitl}}]{Sesar2010}
{Sesar}, B., {Ivezi{\'c}}, {\v Z}., {Grammer}, S.~H., {et~al.} 2010, \apj, 708,
  717

\bibitem[{{Shu} {et~al.}(1994){Shu}, {Najita}, {Ruden}, \& {Lizano}}]{Shu1994}
{Shu}, F.~H., {Najita}, J., {Ruden}, S.~P., \& {Lizano}, S. 1994, \apj, 429,
  797

\bibitem[{{Skrutskie} {et~al.}(2006){Skrutskie}, {Cutri}, {Stiening},
  {Weinberg}, {Schneider}, {Carpenter}, {Beichman}, {Capps}, {Chester},
  {Elias}, {Huchra}, {Liebert}, {Lonsdale}, {Monet}, {Price}, {Seitzer},
  {Jarrett}, {Kirkpatrick}, {Gizis}, {Howard}, {Evans}, {Fowler}, {Fullmer},
  {Hurt}, {Light}, {Kopan}, {Marsh}, {McCallon}, {Tam}, {Van Dyk}, \&
  {Wheelock}}]{Skrutskie2006}
{Skrutskie}, M.~F., {Cutri}, R.~M., {Stiening}, R., {et~al.} 2006, \aj, 131,
  1163

\bibitem[{{Smee} {et~al.}(2013){Smee}, {Gunn}, {Uomoto}, {Roe}, {Schlegel},
  {Rockosi}, {Carr}, {Leger}, {Dawson}, {Olmstead}, {Brinkmann}, {Owen},
  {Barkhouser}, {Honscheid}, {Harding}, {Long}, {Lupton}, {Loomis}, {Anderson},
  {Annis}, {Bernardi}, {Bhardwaj}, {Bizyaev}, {Bolton}, {Brewington}, {Briggs},
  {Burles}, {Burns}, {Castander}, {Connolly}, {Davenport}, {Ebelke}, {Epps},
  {Feldman}, {Friedman}, {Frieman}, {Heckman}, {Hull}, {Knapp}, {Lawrence},
  {Loveday}, {Mannery}, {Malanushenko}, {Malanushenko}, {Merrelli}, {Muna},
  {Newman}, {Nichol}, {Oravetz}, {Pan}, {Pope}, {Ricketts}, {Shelden},
  {Sandford}, {Siegmund}, {Simmons}, {Smith}, {Snedden}, {Schneider},
  {SubbaRao}, {Tremonti}, {Waddell}, \& {York}}]{Smee2013}
{Smee}, S.~A., {Gunn}, J.~E., {Uomoto}, A., {et~al.} 2013, \aj, 146, 32

\bibitem[{{Stepanov} {et~al.}(2001){Stepanov}, {Kliem}, {Zaitsev}, {F{\"u}rst},
  {Jessner}, {Kr{\"u}ger}, {Hildebrandt}, \& {Schmitt}}]{Stepanov2001}
{Stepanov}, A.~V., {Kliem}, B., {Zaitsev}, V.~V., {et~al.} 2001, \aap, 374,
  1072

\bibitem[{{Stokes} {et~al.}(1998){Stokes}, {Viggh}, {Shelly}, {Blythe}, \&
  {Stuart}}]{Stokes1998}
{Stokes}, G.~H., {Viggh}, H.~E.~M., {Shelly}, F.~L., {Blythe}, M.~S., \&
  {Stuart}, J.~S. 1998, in Bulletin of the American Astronomical Society,
  Vol.~30, AAS/Division for Planetary Sciences Meeting Abstracts \#30, 1042

\bibitem[{{Stoughton}(2002)}]{Stoughton2002b}
{Stoughton}, C. 2002, \aj, 123, 3487

\bibitem[{{Stoughton} {et~al.}(2002){Stoughton}, {Lupton}, {Bernardi},
  {Blanton}, {Burles}, {Castander}, {Connolly}, {Eisenstein}, {Frieman},
  {Hennessy}, {Hindsley}, {Ivezi{\'c}}, {Kent}, {Kunszt}, {Lee}, {Meiksin},
  {Munn}, {Newberg}, {Nichol}, {Nicinski}, {Pier}, {Richards}, {Richmond},
  {Schlegel}, {Smith}, {Strauss}, {SubbaRao}, {Szalay}, {Thakar}, {Tucker},
  {Vanden Berk}, {Yanny}, {Adelman}, {Anderson}, {Anderson}, {Annis},
  {Bahcall}, {Bakken}, {Bartelmann}, {Bastian}, {Bauer}, {Berman},
  {B{\"o}hringer}, {Boroski}, {Bracker}, {Briegel}, {Briggs}, {Brinkmann},
  {Brunner}, {Carey}, {Carr}, {Chen}, {Christian}, {Colestock}, {Crocker},
  {Csabai}, {Czarapata}, {Dalcanton}, {Davidsen}, {Davis}, {Dehnen},
  {Dodelson}, {Doi}, {Dombeck}, {Donahue}, {Ellman}, {Elms}, {Evans}, {Eyer},
  {Fan}, {Federwitz}, {Friedman}, {Fukugita}, {Gal}, {Gillespie}, {Glazebrook},
  {Gray}, {Grebel}, {Greenawalt}, {Greene}, {Gunn}, {de Haas}, {Haiman},
  {Haldeman}, {Hall}, {Hamabe}, {Hansen}, {Harris}, {Harris}, {Harvanek},
  {Hawley}, {Hayes}, {Heckman}, {Helmi}, {Henden}, {Hogan}, {Hogg}, {Holmgren},
  {Holtzman}, {Huang}, {Hull}, {Ichikawa}, {Ichikawa}, {Johnston}, {Kauffmann},
  {Kim}, {Kimball}, {Kinney}, {Klaene}, {Kleinman}, {Klypin}, {Knapp},
  {Korienek}, {Krolik}, {Kron}, {Krzesi{\'n}ski}, {Lamb}, {Leger},
  {Limmongkol}, {Lindenmeyer}, {Long}, {Loomis}, {Loveday}, {MacKinnon},
  {Mannery}, {Mantsch}, {Margon}, {McGehee}, {McKay}, {McLean}, {Menou},
  {Merelli}, {Mo}, {Monet}, {Nakamura}, {Narayanan}, {Nash}, {Neilsen},
  {Newman}, {Nitta}, {Odenkirchen}, {Okada}, {Okamura}, {Ostriker}, {Owen},
  {Pauls}, {Peoples}, {Peterson}, {Petravick}, {Pope}, {Pordes}, {Postman},
  {Prosapio}, {Quinn}, {Rechenmacher}, {Rivetta}, {Rix}, {Rockosi}, {Rosner},
  {Ruthmansdorfer}, {Sandford}, {Schneider}, {Scranton}, {Sekiguchi}, {Sergey},
  {Sheth}, {Shimasaku}, {Smee}, {Snedden}, {Stebbins}, {Stubbs}, {Szapudi},
  {Szkody}, {Szokoly}, {Tabachnik}, {Tsvetanov}, {Uomoto}, {Vogeley}, {Voges},
  {Waddell}, {Walterbos}, {Wang}, {Watanabe}, {Weinberg}, {White}, {White},
  {Wilhite}, {Wolfe}, {Yasuda}, {York}, {Zehavi}, \& {Zheng}}]{Stoughton2002}
{Stoughton}, C., {Lupton}, R.~H., {Bernardi}, M., {et~al.} 2002, \aj, 123, 485

\bibitem[{{S{\"u}veges} {et~al.}(2012){S{\"u}veges}, {Sesar}, {V{\'a}radi},
  {Mowlavi}, {Becker}, {Ivezi{\'c}}, {Beck}, {Nienartowicz}, {Rimoldini},
  {Dubath}, {Bartholdi}, \& {Eyer}}]{Suveges2012}
{S{\"u}veges}, M., {Sesar}, B., {V{\'a}radi}, M., {et~al.} 2012, \mnras, 424,
  2528

\bibitem[{{Theissen} {et~al.}(2015){Theissen}, {West}, \&
  {Dhital}}]{Theissen2015}
{Theissen}, C.~A., {West}, A.~A., \& {Dhital}, S. 2015, ArXiv e-prints,
  arXiv:1509.01907

\bibitem[{{Walkowicz} {et~al.}(2011){Walkowicz}, {Basri}, {Batalha},
  {Gilliland}, {Jenkins}, {Borucki}, {Koch}, {Caldwell}, {Dupree}, {Latham},
  {Meibom}, {Howell}, {Brown}, \& {Bryson}}]{Walkowicz2011}
{Walkowicz}, L.~M., {Basri}, G., {Batalha}, N., {et~al.} 2011, \aj, 141, 50

\bibitem[{{Ward-Duong} {et~al.}(2015){Ward-Duong}, {Patience}, {De Rosa},
  {Bulger}, {Rajan}, {Goodwin}, {Parker}, {McCarthy}, \&
  {Kulesa}}]{Ward-Duong2015}
{Ward-Duong}, K., {Patience}, J., {De Rosa}, R.~J., {et~al.} 2015, \mnras, 449,
  2618

\bibitem[{{Watkins} {et~al.}(2009){Watkins}, {Evans}, {Belokurov}, {Smith},
  {Hewett}, {Bramich}, {Gilmore}, {Irwin}, {Vidrih}, {Wyrzykowski}, \&
  {Zucker}}]{Watkins2009}
{Watkins}, L.~L., {Evans}, N.~W., {Belokurov}, V., {et~al.} 2009, \mnras, 398,
  1757

\bibitem[{{Webbink}(1988)}]{Webbink1988}
{Webbink}, R.~F. 1988, {The Formation and Evolution of Symbiotic Stars}, ed.
  J.~{Mikolajewska}, M.~{Friedjung}, S.~J. {Kenyon}, \& R.~{Viotti}, 311

\bibitem[{{Welsh} {et~al.}(2012){Welsh}, {Orosz}, {Carter}, {Fabrycky}, {Ford},
  {Lissauer}, {Pr{\v s}a}, {Quinn}, {Ragozzine}, {Short}, {Torres}, {Winn},
  {Doyle}, {Barclay}, {Batalha}, {Bloemen}, {Brugamyer}, {Buchhave},
  {Caldwell}, {Caldwell}, {Christiansen}, {Ciardi}, {Cochran}, {Endl},
  {Fortney}, {Gautier}, {Gilliland}, {Haas}, {Hall}, {Holman}, {Howard},
  {Howell}, {Isaacson}, {Jenkins}, {Klaus}, {Latham}, {Li}, {Marcy}, {Mazeh},
  {Quintana}, {Robertson}, {Shporer}, {Steffen}, {Windmiller}, {Koch}, \&
  {Borucki}}]{Welsh2012}
{Welsh}, W.~F., {Orosz}, J.~A., {Carter}, J.~A., {et~al.} 2012, \nat, 481, 475

\bibitem[{{West} {et~al.}(2008){West}, {Hawley}, {Bochanski}, {Covey}, {Reid},
  {Dhital}, {Hilton}, \& {Masuda}}]{West2008}
{West}, A.~A., {Hawley}, S.~L., {Bochanski}, J.~J., {et~al.} 2008, \aj, 135,
  785

\bibitem[{{West} {et~al.}(2004){West}, {Hawley}, {Walkowicz}, {Covey},
  {Silvestri}, {Raymond}, {Harris}, {Munn}, {McGehee}, {Ivezi{\'c}}, \&
  {Brinkmann}}]{West2004}
{West}, A.~A., {Hawley}, S.~L., {Walkowicz}, L.~M., {et~al.} 2004, \aj, 128,
  426

\bibitem[{{West} {et~al.}(2011){West}, {Morgan}, {Bochanski}, {Andersen},
  {Bell}, {Kowalski}, {Davenport}, {Hawley}, {Schmidt}, {Bernat}, {Hilton},
  {Muirhead}, {Covey}, {Rojas-Ayala}, {Schlawin}, {Gooding}, {Schluns},
  {Dhital}, {Pineda}, \& {Jones}}]{West2011}
{West}, A.~A., {Morgan}, D.~P., {Bochanski}, J.~J., {et~al.} 2011, \aj, 141, 97

\bibitem[{{Winters} {et~al.}(2015){Winters}, {Henry}, {Lurie}, {Hambly}, {Jao},
  {Bartlett}, {Boyd}, {Dieterich}, {Finch}, {Hosey}, {Ianna}, {Riedel},
  {Slatten}, \& {Subasavage}}]{Winters2015}
{Winters}, J.~G., {Henry}, T.~J., {Lurie}, J.~C., {et~al.} 2015, \aj, 149, 5

\bibitem[{{York} {et~al.}(2000){York}, {Adelman}, {Anderson}, {Anderson},
  {Annis}, {Bahcall}, {Bakken}, {Barkhouser}, {Bastian}, {Berman}, {Boroski},
  {Bracker}, {Briegel}, {Briggs}, {Brinkmann}, {Brunner}, {Burles}, {Carey},
  {Carr}, {Castander}, {Chen}, {Colestock}, {Connolly}, {Crocker}, {Csabai},
  {Czarapata}, {Davis}, {Doi}, {Dombeck}, {Eisenstein}, {Ellman}, {Elms},
  {Evans}, {Fan}, {Federwitz}, {Fiscelli}, {Friedman}, {Frieman}, {Fukugita},
  {Gillespie}, {Gunn}, {Gurbani}, {de Haas}, {Haldeman}, {Harris}, {Hayes},
  {Heckman}, {Hennessy}, {Hindsley}, {Holm}, {Holmgren}, {Huang}, {Hull},
  {Husby}, {Ichikawa}, {Ichikawa}, {Ivezi{\'c}}, {Kent}, {Kim}, {Kinney},
  {Klaene}, {Kleinman}, {Kleinman}, {Knapp}, {Korienek}, {Kron}, {Kunszt},
  {Lamb}, {Lee}, {Leger}, {Limmongkol}, {Lindenmeyer}, {Long}, {Loomis},
  {Loveday}, {Lucinio}, {Lupton}, {MacKinnon}, {Mannery}, {Mantsch}, {Margon},
  {McGehee}, {McKay}, {Meiksin}, {Merelli}, {Monet}, {Munn}, {Narayanan},
  {Nash}, {Neilsen}, {Neswold}, {Newberg}, {Nichol}, {Nicinski}, {Nonino},
  {Okada}, {Okamura}, {Ostriker}, {Owen}, {Pauls}, {Peoples}, {Peterson},
  {Petravick}, {Pier}, {Pope}, {Pordes}, {Prosapio}, {Rechenmacher}, {Quinn},
  {Richards}, {Richmond}, {Rivetta}, {Rockosi}, {Ruthmansdorfer}, {Sandford},
  {Schlegel}, {Schneider}, {Sekiguchi}, {Sergey}, {Shimasaku}, {Siegmund},
  {Smee}, {Smith}, {Snedden}, {Stone}, {Stoughton}, {Strauss}, {Stubbs},
  {SubbaRao}, {Szalay}, {Szapudi}, {Szokoly}, {Thakar}, {Tremonti}, {Tucker},
  {Uomoto}, {Vanden Berk}, {Vogeley}, {Waddell}, {Wang}, {Watanabe},
  {Weinberg}, {Yanny}, {Yasuda}, \& {SDSS Collaboration}}]{York2000}
{York}, D.~G., {Adelman}, J., {Anderson}, Jr., J.~E., {et~al.} 2000, \aj, 120,
  1579

\bibitem[{{Zhilyaev} {et~al.}(2007){Zhilyaev}, {Romanyuk}, {Svyatogorov},
  {Verlyuk}, {Kaminsky}, {Andreev}, {Sergeev}, {Gershberg}, {Lovkaya},
  {Avgoloupis}, {Seiradakis}, {Contadakis}, {Antov}, {Konstantinova-Antova}, \&
  {Bogdanovski}}]{Zhilyaev2007}
{Zhilyaev}, B.~E., {Romanyuk}, Y.~O., {Svyatogorov}, O.~A., {et~al.} 2007,
  \aap, 465, 235

\end{thebibliography}
\bibliographystyle{apj}

\end{document}